\documentclass[journal,article,accept,moreauthors]{Definitions/mdpi} 

\DeclareUnicodeCharacter{02BE}{\textquotesingle}
\DeclareUnicodeCharacter{1E93}{\d{z}}
\DeclareUnicodeCharacter{02BF}{\textquotesingle}
\DeclareUnicodeCharacter{1E6C}{\d{T}}
\DeclareUnicodeCharacter{1E62}{\d{S}}
\DeclareUnicodeCharacter{1E24}{\d{H}}
\DeclareUnicodeCharacter{0331}{\_}
\DeclareUnicodeCharacter{1E95}{\b{z}}
\DeclareUnicodeCharacter{1E92}{\d{Z}}
\DeclareUnicodeCharacter{035F}{\b{}}

\usepackage{algorithm}
\usepackage{algpseudocode}
\usepackage[utf8]{inputenc}
\usepackage[arabic,english]{babel}
\usepackage{tipa}
\usepackage{longtable}

\firstpage{1} 
\pubvolume{1}
\issuenum{1}
\articlenumber{1}
\pubyear{2025}
\copyrightyear{2025}
\datereceived{ } 
\daterevised{ } 
\dateaccepted{ } 
\datepublished{ } 
\hreflink{https://doi.org/} 
\pdfoutput=1 

\Title{NAZM: Network Analysis of Zonal Metrics in Persian Poetic Tradition}

\TitleCitation{Title}

\Author{
Kourosh Shahnazari $^{1,\dagger}$ \orcidA{} , Seyed Moein Ayyoubzadeh  $^{1,\dagger}$ \orcidB{} , Mohammadamin Fazli  $^{1}$ \orcidC{}  , Mohammadali Keshtparvar $^{2}$ \orcidD{}  }

\AuthorNames{}

\isAPAStyle{%
       \AuthorCitation{Lastname, F., Lastname, F., \& Lastname, F.}
         }{%
        \isChicagoStyle{%
        \AuthorCitation{Lastname, Firstname, Firstname Lastname, and Firstname Lastname.}
        }{
        \AuthorCitation{Lastname, F.; Lastname, F.; Lastname, F.}
        }
}

\address{$^{1}$ \quad Sharif University of Technology\\$^{2}$ \quad Amirkabir University of Technology
}

\corres{}

\firstnote{These authors contributed equally to this work.
}

\abstract{This study formalizes a computational model to simulate classical Persian poets' dynamics of influence through constructing a multi-dimensional similarity network. Using a rigorously curated dataset based on Ganjoor's corpus, we draw upon semantic, lexical, stylistic, thematic, and metrical features to demarcate each poet's corpus. Each is contained within weighted similarity matrices, which are then appended to generate an aggregate graph showing poet-to-poet influence. Further network investigation is carried out to identify key poets, style hubs, and bridging poets by calculating degree, closeness, betweenness, eigenvector, and Katz centrality measures. Further, for typological insight, we use the Louvain community detection algorithm to demarcate clusters of poets sharing both style and theme coherence, which correspond closely to acknowledged schools of literature like Sabk-e Hindi, Sabk-e Khorasani, and the Bazgasht-e Adabi phenomenon. Our findings provide a new data-driven view of Persian literature distinguished between canonical significance and interextual influence, thus highlighting relatively lesser-known figures who hold great structural significance. Combining computational linguistics with literary study, this paper produces an interpretable and scalable model for poetic tradition, enabling retrospective reflection as well as forward-looking research within digital humanities.}

\fancypagestyle{plain}{
  \fancyhf{} 
}

\begin{document}


\section{Introduction}

\subsection{Background and Motivation}
For many centuries, Persian poetry has been the means for reflection on philosophy, culture and social criticism. Great poets, such as Ferdowsi, Rumi, Hafez, Saadi, and others, have tremendously enriched the literary heritage, affecting the works of the next poetical generations. Such literary influences were traditionally understood from an angle, held by the judgment of every scholar, through a qualitative textual examination with an investigation of historical contexts and thematic categorization. \citep{asgari2013persian, rahgozar2019hafez}.

Advances in Natural Language Processing and computational linguistics make it possible to quantify the degree of influence by analyzing systematic textual features such as lexical choice, poetic meter, thematic coherence, and syntactic structure. In continuation of earlier work, this paper presents systematic computational analysis to set up an influence graph with the eventual end of building and establishing a dynamic network of poets based on the earlier works in analyzing Persian poetry \citep{meymandi2024ai}.

\subsection{Challenges in Measuring Literary Influence}
Computers may be useful, but literary influence has numerous challenges to analyze:

\noindent
\textbf{1. Knowing References and Effects:} 
All forms of textual similarity cannot be carried influence; poets might cite earlier works as homage, use speech acts for rhetorical ends, or oblique allusions. Open influence is often much harder to quantify than subconscious influence \citep{zhao2022socialnet, rezaei2017stylometric}.

\noindent
\textbf{2. Diachronic Linguistic Evolution:} 
Persian poetry, which has been in existence for more than a thousand years now, has seen considerable linguistic changes: the old pre-Islamic vocabulary, the Arabic borrowings during the Islamic period, and the refinements of modern Persian. The channels of influence exerted upon poets and poetry are, nevertheless, sometimes recognized only through hard lexical normalizations \citep{asgari2013persian}.

\noindent
\textbf{3. Poetic Form and Meter Complexity} Classical Persian poetry is governed by certain meters called "Arud," affecting the choice of style. Special computer techniques are needed to separate these rules of rhythm from their influence on style \citep{rahgozar2020semantic}.

\noindent
\textbf{4. Centrality Bias and Graph Density within Network Analysis:} 
Excessive algorithmic bias may render certain critical links impassable in the larger groups of poetry. Abstracting from the present scenario, data-population models with high centrality measures may actually be indicative of how nicely the set is created rather than fielding current patterns of influence \citep{milonia2023temporal}.

\subsection{Computational Framework for Influence Analysis}
\label{subsec:influence-framework}

Traditionally, studies of literary influence have been primarily concerned with close reading, historiography, and critical interpretation, but the search for structured methodologies to follow the transmission of themes, styles, and intellectual paradigms credited to their lineage among generations of poets has long been undertaken. With computational humanities and novel developments in natural language processing (NLP), it has now become possible to treat literary influence as an operationalized, measurable, and analyzable phenomenon. In this paper, we present a unified computational framework for the analysis of influence among classical Persian poets, utilizing multi-level similarity measures, graph-theoretical formulations, and centrality analysis to model, quantify, and interpret inter-poet relationships that are scalable and reproducible.

Our approach begins with the collection and preprocessing of a large corpus of Persian poetry, encompassing multiple poets spanning different historical periods, literary schools, and stylistic orientations. For each poet, we consolidate their poetic output into a unified representation suitable for linguistic analysis. These representations serve as the foundation for a multi-dimensional similarity analysis, wherein we evaluate how closely poets align across several axes of comparison.

At the core of our methodology are the construction of the five similarity matrices corresponding to different literary dimensions: \textit{semantic}, \textit{stylistic}, \textit{thematic}, \textit{metrical}, and \textit{lexical}. Semantic similarity measures meaning-based relationships between poets by performing a comparison using Word2Vec word-level embeddings or frequency-weighted matching algorithms. Stylistic similarity is concerned with formal and grammatical patterns: certain statistical features are extracted, mainly those of verse length, lexical diversity, and distribution of parts of speech. Thematic similarity is derived from deep sentence embeddings from transformer models capable of capturing high-level subjects and motifs of poetry. Meter similarity is oriented toward structural and rhythmic tendencies, utilizing bag-of-pattern instances over annotations of poetic meter. Lexical similarity, lastly, evaluates direct overlap in vocabulary and expression through TF-IDF vectorization and cosine similarity. Each of these dimensions produces a fully connected, weighted poet-poet similarity graph, bringing new perspectives on how influence may be transmitted and shared.

To build a computationally tractable and interpretable graph of influence, we integrate these similarity measures into a fused matrix and apply filtering techniques to retain only the strongest and most meaningful connections. For each poet, we impose a minimum similarity threshold to avoid noise and spurious edges. This process yields a sparse, interpretable, and empirically grounded influence graph, where nodes represent poets and weighted edges signify high multi-dimensional similarity—serving as proxies for literary influence.

Once the influence graph has been constructed, the analysis will lead to the inquiry into its structural characteristics. Exploiting terms from network science, we calculate a variety of centrality measures for each poet: degree centrality that captures the breadth of direct influence; betweenness centrality showing bridge roles in the flow of influence; closeness centrality measures for how fast a poet can access the rest of the network; eigenvector centrality rewarding connection to other influential poets; and Katz centrality, allowing influence to be spread through indirect means. Those metrics allow the formal identification of the most influential, well-connected, and structurally pivotal poets in the graph. Furthermore, community detection algorithms may be applied to outline clusters of poetic schools or groups of stylistic affinities, complementing the interpretability of the network.

Computational operations are entirely automatable and reproducible; that is, from computing similarity through the construction of the graph to finally estimating the centrality thereof. The method we present is one that is scalable, which grounds our influence analysis in quantifiable and explainable linguistic features, offering a possible linkage between computational methods and literary theory. Most important, our approach neither replaces nor replaces the humanistic interpretation but assists it with empirical tools that lend support, challenge, or refine existing narratives regarding influence in Persian literary historiography.

This framework represents a significant advancement in digital literary studies of Persian poetry. It not only enables large-scale, fine-grained comparison of poets across multiple dimensions, but also opens the door to further analyses such as temporal evolution of influence, stylistic diffusion modeling, and authorship disambiguation. As such, it contributes to the growing field of computational poetics and offers a template for influence modeling in other literary traditions and languages.

\subsection{Expected Insights and Future Directions}
\label{subsec:expected-insights}

By constructing a multi-dimensional influence graph of classical Persian poets, this investigation aspires to produce a range of new and quantifiable insights into the structure and dynamics of Persian literary heritage. Among the major goals is the identification of poets in the network whose positions are pivotal—centers of stylistic innovation, semantic bridge-builders, or otherwise thematic focal points. We think that through combining centrality measures across five orthogonal similarity spaces - namely, semantic, stylistic, thematic, metrical, and lexical-we will find poets who exercised wide-ranging influence in multiple spheres of poetic expression. Those poets historically may or may not be celebrated, thus providing a framework for revise literary canons and surface under-celebrated contributors whose influence has remained unwarranted in traditional scholarship.

The structural topology of the graphs might be seized upon to discern literary communities with distinct meanings. If we were to apply some form of unsupervised community detection to possible clusters of poets aligned stylistically and thematically, we believe we might discover a number of clusters that might correspond to recognized literary schools, time periods, and geographical regions, or perhaps even newly envisioned groupings not yet acknowledged. Thus, this framework provides the data-driven basis for querying literary genealogy and intellectual lineage.

The methodology also sets the foundation for several promising future research directions. Temporal modeling can be implemented by superimposing chronological metadata onto the influence graph, thus enabling diachronic analysis of the evolution, divergence, and convergence of poetic styles over time. In this way, it can aid the tracing of stylistic drift, the emergence of thematic paradigms, and the life cycle of metrical traditions. The second prospective extension of this framework is classifying it as the task of \textit{disputed authorship resolution} in which the similarity profiles of anonymous or contested poems are compared with the profiles of known poets to infer potential author. Thirdly, one would use more sophisticated neural embedding techniques, such as fine-tuning pre-trained transformer models specifically for Persian verse, in order to improve the quality and domain specificity of the embeddings made use of in semantic and thematic similarity calculations.

Another key direction involves the interpretability and visualization of influence networks. By integrating interactive graph visualization platforms (e.g., Plotly Dash or D3.js), researchers, educators, and students could explore the structure of Persian literary history dynamically, facilitating humanistic inquiry augmented by data-driven tools. Additionally, the approach is easily generalizable and can be replicated across different literary corpora, languages, and genres, contributing to the broader field of computational comparative literature.

Finally, by making all code, datasets, and models publicly available, we aim to foster reproducibility, transparency, and collaborative growth in Persian digital humanities. This project thus not only contributes new insights to Persian literary studies, but also advances the methodological toolkit available to researchers engaging with literature at scale.

\section{Related Work}
\subsection{Computational Poetry Analysis}

Computational poetry analysis seeks to extract quantitative insights from poetic texts by examining lexical patterns, metrical structures, thematic consistencies, and stylistic similarities. Early efforts in this domain primarily employed statistical and rule-based methods to identify rhyme schemes and meter patterns, providing foundational insights into the formal aspects of poetry. For example, \citet{greene2010automatic} utilized algorithmic approaches to detect rhyme schemes and metrical patterns in English poetry, laying the groundwork for quantitative analysis of poetic structures.

The introduction of machine learning techniques has significantly advanced the field, enabling more nuanced analysis of poetic texts. One pioneering approach is the use of word embeddings like Word2Vec and FastText, which map words into high-dimensional vector spaces based on their contextual usage \citep{mikolov2013efficient}. These embeddings have proven particularly useful for analyzing lexical similarity and detecting thematic consistencies between poets. For instance, \citet{iyyer2014neural} applied neural embeddings to literary texts to capture latent stylistic connections, revealing previously unrecognized influences between authors.

Another widely used technique in computational poetry analysis is the Term Frequency-Inverse Document Frequency (TF-IDF) method \citep{salton1975vector}. TF-IDF measures the importance of words within a document relative to a collection of documents (corpus). It has been effectively employed to identify characteristic keywords and lexical patterns within poetic texts. By analyzing the TF-IDF scores of words across multiple poets, researchers can detect distinctive lexical choices and thematic preferences. This technique has been used in combination with clustering algorithms to group poets based on lexical similarity, providing insights into shared vocabulary and stylistic conventions.

The emergence of transformer-based models has further revolutionized computational poetry analysis. The BERT model (Bidirectional Encoder Representations from Transformers) has significantly enhanced the understanding of semantic relationships within poetic texts by modeling contextual meaning and capturing subtle semantic shifts \citep{devlin2019bert}. Following BERT, RoBERTa (Robustly optimized BERT approach) introduced several optimizations that improved model performance, such as dynamic masking and training on larger corpora \citep{liu2019roberta}. RoBERTa has demonstrated superior capabilities in capturing contextual nuances and has been successfully applied in analyzing more complex literary structures, including metaphorical language and nuanced stylistic variations. Its enhanced training dynamics make it particularly suitable for dealing with rich and complex poetic texts, where subtle differences in word usage and syntax carry significant interpretative weight.

Despite these advances, the application of computational poetry analysis to classical Persian literature remains relatively unexplored. Most existing research focuses on modern texts or Western literary traditions, neglecting the unique linguistic and stylistic features of Persian poetry. Our study addresses this gap by incorporating a multi-dimensional approach that combines lexical, metrical, thematic, citation-based, and stylistic analyses to map the intricate web of poetic influence among classical Persian poets.

\subsection{Graph-Based Influence Detection}

Analyzing influence and intertextuality in literary studies has become mostly dependent on graph-based approaches. These approaches help to visualize and count complicated relationships among literary figures by presenting poets as nodes and their influences as weighted edges. Graph theory and network analysis methods help to find central figures, groups of stylistically similar poets, and links spanning geographical and chronological limits.

One prominent example is the research by \citet{zhao2022socialnet}, which created a thorough social network of ancient Chinese writers to investigate the change of their impact over time. To create a network of 41,310 poets, this study combined multisourced data including biographies of poets and lyrical works. Using propagation dynamics models to assess influence distribution and a time-series entropy weight method to measure dynamic changes in poets' influence over time, the authors presented This creative method allowed the study of literary development from a data-driven standpoint and the identification of significant writers. Natural language processing and social network analysis approaches combined in the study gave important new perspectives on the relationships between poets from several dynasties and schools.

Graph-based approaches have been rather underused in Persian poetry even if they have been successful in other literary traditions. Many times, existing research concentrate on stylistic or thematic analysis without clearly referencing poet-to--poet inspirations. Our method combines several aspects of similarity—lexical, metrical, thematic, citation-based, and stylistic—into a unified influence graph in order to close this gap For classical Persian poets, this all-encompassing model improves the identification of subtle influences and thematic continuity.

Moreover, we use cutting-edge graph analysis methods to compute centrality measures, clustering coefficients, and community structures, so enabling the identification of important poets and literary movements inside the Persian poetic tradition. Additionally included are dynamic graph visualizing methods that allow interactive investigation of influence patterns. Users of Dash and Plotly can quickly negotiate the poetic network while seeing how influences change across poetic genres and over time. This dynamic portrayal not only improves user involvement but also offers closer understanding of the literary development of Persian poetry.


\section{Methodology}

\subsection{Data Collection and Preprocessing}
\subsection*{Dataset Compilation}
This work made use of a main dataset derived from the \textit{Ganjoor}  digital library, an extensive and freely available collection of classical Persian poetry. \citet{ganjoor} Along with related metadata including poetic form, meter, and authorship information, Ganjoor offers well chosen collections of works by hundreds of Persian poets. We compiled each poet's complete set of recorded poems, together with their name, birth year (as noted on the Hijri calendar), and overall count of verses (\textit{beyts}) credited to them. Chronological studies were made easier and historical consistency with Persian literary timelines preserved using the Hijri calendar. Further preprocessing of the dataset helped to eliminate duplicates, normalize textual artifacts, and guarantee consistency in author identification over all obtained data.

\subsubsection{Data Cleaning and Filtering}
Poets with less than 500 lines were not included in order to guarantee the strength of the study. This filtering process sought to reduce the impact of statistical noise from writers with low textual output. Moreover, the dataset was carefully cleaned to solve poet name inconsistencies.

\subsubsection{Text Normalization}
\label{subsec:normalization}

Persian text offers a range of preprocessing difficulties because of variations in spelling rules, character encoding standards, and the existence of orthographic variants passed on from Arabic script. We used the extensively used Persian natural language processing tool \textit{Hazm} library to apply a thorough normalizing process guaranteeing consistency and lowering of noise in the input data. \citet{hazm}

The normalization pipeline consisted of several steps:

\begin{itemize}
    \item \textbf{Standardization of character forms}: Standard Persian forms replaced variants of letters often found in Arabic-script sources. This stage guarantees consistency in the way often occurring words are represented across several texts and prevents mismatches in token-based comparisons.

    \item \textbf{Removal of diacritical marks}: Short vowels and other pronunciation aids were eliminated from the book since they are often inconsistently used in digital corpora and have little bearing on lexical or semantic analysis.

    \item \textbf{Elimination of control characters}: Invisible formatting elements such as zero-width non-joiners and other non-printing Unicode characters were removed to prevent irregular tokenization and improve alignment in similarity measures.

    \item \textbf{Whitespace and punctuation normalization}: Irregular spacing, redundant punctuation, and inconsistent use of separators were corrected to ensure clean and uniform token boundaries.
\end{itemize}

Harmonizing the corpus and guaranteeing that surface-level textual variation did not affect downstream similarity computations depended on this normalizing process. It was fundamental in helping the data be ready for precise lexical, semantic, and stylistic comparisons among poets.

\subsection{Similarity Measures}
\label{sec:similarity}

We define a set of five unique similarity measures, each intended to capture a different dimension of poetic resonance: semantic, stylistic, thematic, metrical, and lexical, so building an influence graph among classical Persian poets. These steps provide the weighted edges of a poet-poet similarity graph, so allowing a thorough multi-view study of proximity and influence in Persian literary tradition.

\subsubsection{Semantic Similarity}
\label{subsec:semantic}

Semantic similarity seeks to measure between poets the alignment of conceptual and contextual word use. We obtained dense vector embeddings for individual words by using a Word2Vec model trained on Persian literary corpora. We created a frequency-weighted vocabulary list for every poet whereby every word mapped to its matching embedding vector.

Using a maximum-based cosine similarity approach, the similarity between two poets was computed: the mean of these maxima was obtained by computing, for every word vector in the vocabulary of poet $A$, the maximum cosine similarity to any word vector in the vocabulary of poet $B$. This procedure averaged for symmetry and was carried out both directions. To enhance distinctiveness, we applied the following modifications:

\begin{itemize}
    \item \textbf{Frequency similarity}: Penalized imbalanced usage of shared words via squared frequency ratios.
    \item \textbf{Jaccard dissimilarity}: Incorporated word set overlap as a disambiguation filter.
    \item \textbf{Non-linear scaling}: Boosted contrast between weak and strong similarities using exponential transformation.
    \item \textbf{IDF-based weighting}: Applied an inverse document frequency function to suppress common high-frequency terms.
\end{itemize}

This approach generates a strong semantic similarity measure considering lexical distinctiveness and usage frequency, so capturing conceptual closeness.

\subsubsection{Stylistic Similarity}
\label{subsec:style}

Stylistic similarity reflects structural and grammatical writing features of each poet. Using the Hazm library for Persian NLP, we extracted a set of low-level linguistic features from each poet’s corpus, including:

\begin{itemize}
    \item Average mesrāʿ (verse) length and word length
    \item Lexical diversity (type-token ratio)
    \item Part-of-speech (POS) diversity and frequency ratios for nouns, verbs, adjectives, and adverbs
    \item Word length variance and standard deviation
    \item Estimated syllable complexity
\end{itemize}

All feature vectors were normalized using z-score standardization and reduced via \textbf{Principal Component Analysis (PCA)} to eliminate redundancy and emphasize informative patterns. Final similarities were computed through a weighted combination of cosine, Euclidean, and Manhattan distances between stylistic vectors, forming a smooth yet contrastive stylistic similarity matrix.

\subsubsection{Thematic Similarity}
\label{subsec:theme}

To evaluate thematic proximity, we used the \texttt{multilingual-e5-large-instruct} model, a transformer-based multilingual sentence encoder, to embed the semantic content of individual poems. For each poet, all embeddings were averaged to construct a poet-level thematic representation.

We reduced embedding dimensionality to 64 components using PCA, so maintaining dominant thematic variance. Then, between these lowered embeddings, we calculated several similarity measures including cosine, Euclidean, Manhattan, and Bray-Curtis. To enhance sensitivity, we squared each similarity value (non-linear transformation) and fused them through a weighted sum:

\[
\text{Similarity} = 0.5 \cdot \text{CosSim} + 0.2 \cdot \text{Euclidean} + 0.2 \cdot \text{Manhattan} + 0.1 \cdot \text{Bray-Curtis}
\]

The final score captures the poets’ affinity in subject matter, emotional tone, philosophical orientation, and narrative focus.

\subsubsection{Meter Similarity}
\label{subsec:meter}

Metrical restrictions define much of classical Persian poetry. We aggregated the metrical pattern of every poem (as obtained from scansion tools or metadata) into a frequency distribution per poet to get metrical similarity. A bag-of- patterns model was used via \texttt{CountVectorizer} to vectorize these aggregated meter profiles.

We then computed cosine similarity between meter vectors, producing a measure of rhythmic alignment. This similarity reflects shared usage of metrical patterns and prosodic tendencies, contributing to formal influence estimation.

\subsubsection{Lexical Similarity}
\label{subsec:lexical}

Lexical similarity captures overlap in surface-level word usage. We concatenated all cleaned poems by each poet and applied \texttt{TfidfVectorizer} with a vocabulary cap of 100{,}000 features. The resulting TF-IDF vectors encode each poet’s emphasis on specific terms and idiomatic expressions.

Cosine similarity was used to compare these vectors, yielding a direct measure of vocabulary overlap and usage frequency. This dimension reflects stylistic habits, dictional preferences, and literary register.

\subsubsection{Multi-dimensional Integration}
\label{subsec:integration}

Every kind of similarity generates an independent weighted adjacency matrix. Although all five matrices were independently examined to maintain interpretability, an average similarity matrix was also created by aggregating normalized versions of each dimension. The final influence graph and poet centrality calculations were produced from this fused view.

\subsubsection{Calculation of Network Metrics}
\label{subsec:network-metrics}

Following the computation and fusion of the multi-dimensional similarity matrices into a single similarity graph, the next phase was a thorough investigation of the resultant network structure to measure poet influence. Within this framework, every node stands for a classical Persian poet, and weighted edges indicate strong similarity relationships produced from semantic, stylistic, thematic, lexical, and metrical aspects. We computed a set of well-established centrality measures from network theory, each of which catches a different aspect of influence or positional prominence, in order to find the most structurally important and influential poets inside this graph.

\subsubsection*{Degree Centrality}
Degree centrality is one of the most intuitive and fundamental measures of node importance in a graph. For a given poet (node) \( v \), degree centrality \( C_D(v) \) is defined as the number of direct connections it has to other poets:

\[
C_D(v) = \frac{\deg(v)}{n-1}
\]

where \( \deg(v) \) is the degree of node \( v \), and \( n \) is the total number of nodes in the graph. In the context of our analysis, a high degree centrality indicates a poet whose work shares strong similarities with many others, suggesting either widespread influence or thematic/stylistic commonality with the broader poetic tradition.

\subsubsection*{Betweenness Centrality}
Betweenness centrality captures the extent to which a poet acts as a bridge or intermediary in the network. It is defined as the proportion of shortest paths between all pairs of poets that pass through a given poet \( v \):

\[
C_B(v) = \sum_{s \neq v \neq t} \frac{\sigma_{st}(v)}{\sigma_{st}}
\]

Here, \( \sigma_{st} \) is the total number of shortest paths from node \( s \) to node \( t \), and \( \sigma_{st}(v) \) is the number of those paths that pass through \( v \). Poets with high betweenness centrality play key roles in connecting otherwise distant literary communities, facilitating indirect influence across styles, regions, or periods.

\subsubsection*{Closeness Centrality}
Closeness centrality measures how close a poet is to all other poets in the network, in terms of the shortest path distances. It is formally defined as the inverse of the sum of the shortest distances from node \( v \) to all other nodes:

\[
C_C(v) = \frac{n-1}{\sum_{u \neq v} d(v, u)}
\]

where \( d(v, u) \) denotes the shortest path length between nodes \( v \) and \( u \). A high closeness score suggests a poet who is semantically and stylistically accessible to many others, potentially reflecting centrality in the intellectual or cultural space.

\subsubsection*{Eigenvector Centrality}
Eigenvector centrality generalizes degree centrality by incorporating the importance of a node’s neighbors. A poet connected to highly influential poets receives a higher score than one connected to peripheral figures. It is defined as:

\[
C_E(v) = \frac{1}{\lambda} \sum_{u \in N(v)} A_{vu} C_E(u)
\]

where \( A_{vu} \) is the weight of the edge between poets \( v \) and \( u \), \( N(v) \) is the set of neighbors of \( v \), and \( \lambda \) is a normalization constant (the largest eigenvalue of the adjacency matrix). Eigenvector centrality helps identify core figures whose influence is magnified by their proximity to other influential poets.

\subsubsection*{Katz Centrality}
Katz centrality further refines eigenvector centrality by considering both direct and indirect connections, penalizing longer paths with an attenuation factor. It is given by:

\[
C_K(v) = \alpha \sum_{u \in V} A_{vu} C_K(u) + \beta
\]

Here, \( \alpha \) is a damping factor (typically small to ensure convergence), and \( \beta \) is a constant bias term representing baseline influence. Katz centrality is especially useful in identifying poets who may not have immediate high centrality, but are embedded within influential regions of the graph and contribute to the flow of literary influence across indirect chains. \citet{bloch2023centrality}

\subsubsection*{Implementation Details}
The \texttt{NetworkX} Python tool handled all network computations. Retaining comparable poets for every node, subject to a similarity threshold to remove weak or noisy connections, the graph was built. This sparse, undirected, weighted graph preserves only the most important connections and lets centrality measurements represent core structural roles instead of artifacts of complete connectivity.

\subsubsection*{Interpretation and Output}
Every centrality metric was calculated for every poet and kept in a DataFrame arranged orderly. Our influence study is based on these measures, which are presented graphically as well as tabular. Visualizations highlight overlapping and different influence patterns by means of bar charts of the top-5 poets per centrality metric, network diagrams colored by centrality value, and comparative summaries.

Using these centrality measures, we create a multifarious picture of literary influence—identifying not only the most linked or central figures but also those who function as intellectual conduits, thematic hubs, or stylistic innovators inside the larger Persian poetic tradition.

\subsection{Community Structure (Louvain Method)}
\label{subsec:community-structure}

Apart from using centrality analysis to measure personal poet impact, we aimed to expose the latent \textit{community structure} inside the influence network. In this context, literary communities could match historical eras, artistic movements, regional clusters, or schools of thought; finding them provides insightful analysis of the developing topology of poetic relationships.

We used the widely used and computationally effective \textbf{Louvain method} for community detection to find these communities, a method for which modular structure in big networks can be uncovered. Aiming to maximize a quantity known as \textit{modularity}, a metric of the density of edges inside communities as compared to edges between communities, the Louvain method is a hierarchical, greedy optimization algorithm. \cite{blondel2008louvain} \citet{ramezani2018community}

\subsubsection*{Modularity Maximization}

The modularity \( Q \) of a given partition of a graph is defined as:

\[
Q = \frac{1}{2m} \sum_{i,j} \left[ A_{ij} - \frac{k_i k_j}{2m} \right] \delta(c_i, c_j)
\]

where:
\begin{itemize}
    \item \( A_{ij} \) is the weight of the edge between nodes \( i \) and \( j \),
    \item \( k_i \) and \( k_j \) are the degrees of nodes \( i \) and \( j \),
    \item \( m \) is the total weight of all edges in the network,
    \item \( c_i \) is the community assignment of node \( i \), and
    \item \( \delta(c_i, c_j) \) is the Kronecker delta, which equals 1 if \( c_i = c_j \) and 0 otherwise.
\end{itemize}

A high modularity score indicates a strong community structure: dense connections within communities and sparse connections between them.

\begin{algorithm}[H]
\caption{Louvain Community Detection Algorithm}
\label{alg:louvain}
\begin{algorithmic}[1]
\State \textbf{Input:} Weighted undirected graph $G = (V, E)$
\State \textbf{Output:} Partition of $V$ into communities maximizing modularity

\vspace{0.5em}
\State \textbf{Phase 1: Local Modularity Optimization}
\ForAll{nodes $v \in V$}
    \State Assign $v$ to its own community
\EndFor
\Repeat
    \State improvement $\gets$ \textbf{False}
    \ForAll{nodes $v \in V$}
        \State Let $C_{\text{best}} \gets$ current community of $v$
        \State Let $\Delta Q_{\text{max}} \gets 0$
        \ForAll{neighboring communities $C$ of $v$}
            \State Compute change in modularity $\Delta Q$ if $v$ is moved to $C$
            \If{$\Delta Q > \Delta Q_{\text{max}}$}
                \State $\Delta Q_{\text{max}} \gets \Delta Q$
                \State $C_{\text{best}} \gets C$
            \EndIf
        \EndFor
        \If{$C_{\text{best}} \neq$ current community of $v$}
            \State Move $v$ to $C_{\text{best}}$
            \State improvement $\gets$ \textbf{True}
        \EndIf
    \EndFor
\Until{improvement is \textbf{False}}

\vspace{0.5em}
\State \textbf{Phase 2: Community Aggregation}
\State Build a new graph $G' = (V', E')$ where:
\begin{itemize}
    \item Each node in $V'$ represents a community from $G$
    \item Edge weights in $E'$ represent total edge weights between communities in $G$
\end{itemize}
\State \textbf{If} $G'$ is different from $G$, repeat Phase 1 on $G'$
\State \textbf{Return} final community assignments
\end{algorithmic}
\end{algorithm}

\subsubsection*{Application to Poet Graph}

We applied the Louvain algorithm to our undirected, weighted influence graph using the \texttt{community} package in Python. The similarity weights—derived from the fused similarity matrix—served as edge weights, enabling the detection of semantically and stylistically coherent poetic communities. The resulting community assignments were saved for each poet and used to:

\begin{itemize}
        \item Visualize the community structure in the influence graph via color-coded clusters.
    \item Compute intra- and inter-community centrality averages for comparative analysis.
    \item Support historical or stylistic interpretation of each community's literary characteristics.
\end{itemize}

\subsubsection*{Interpretive Potential}

Community detection complements centrality analysis by revealing the meso-scale organization of the literary landscape. Whereas centrality identifies prominent individuals, community structure uncovers emergent groupings and literary alliances that may reflect shared themes, forms, ideologies, or temporal proximities. By aligning these algorithmic communities with known literary schools or movements (e.g., Khorasani style, Indian style, Isfahani school), we gain both validation and new perspectives on the structural contours of Persian poetic tradition.

\section{Results}
\subsection{Overview of Centrality Rankings}
\label{subsec:centrality-overview}

To quantify the structural importance and network-based influence of each poet in the similarity graph, we computed five centrality metrics: \textit{Degree}, \textit{Betweenness}, \textit{Closeness}, \textit{Eigenvector}, and \textit{Katz} centrality. Each of these metrics captures a distinct aspect of prominence within the graph and helps surface poets who may have served as central figures in the semantic, stylistic, and thematic evolution of Persian poetry.

\begin{figure}[htbp]
    \centering
    \includegraphics[width=0.45\textwidth]{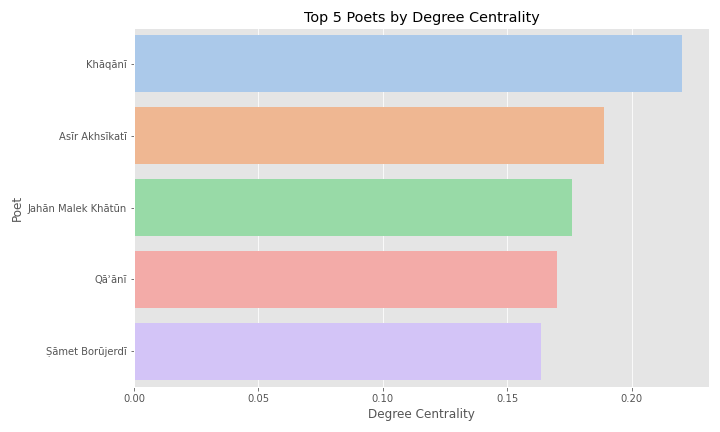}
    \includegraphics[width=0.45\textwidth]{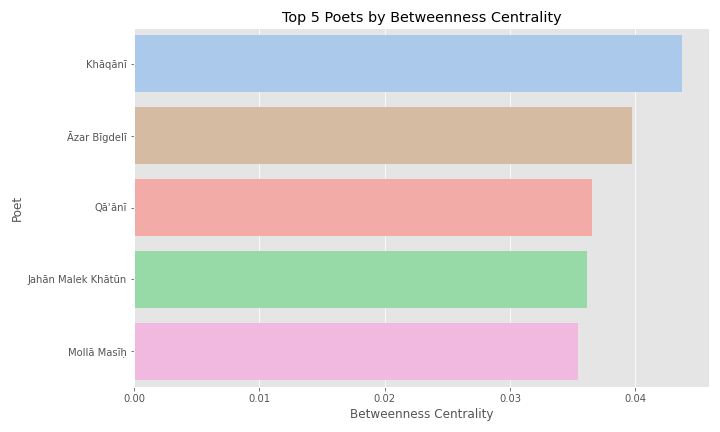}
    
    \includegraphics[width=0.45\textwidth]{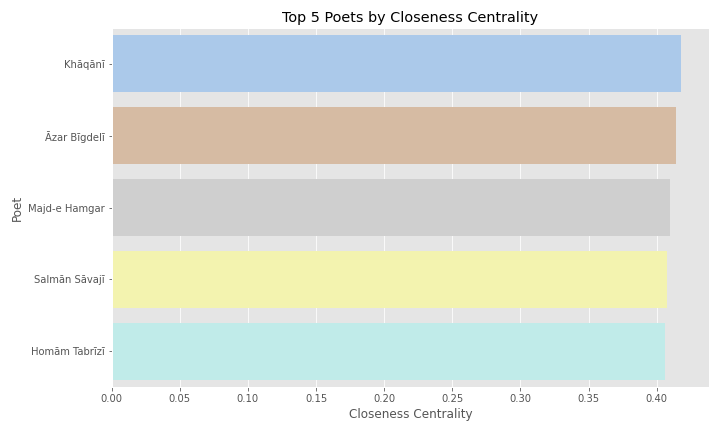}
    \includegraphics[width=0.45\textwidth]{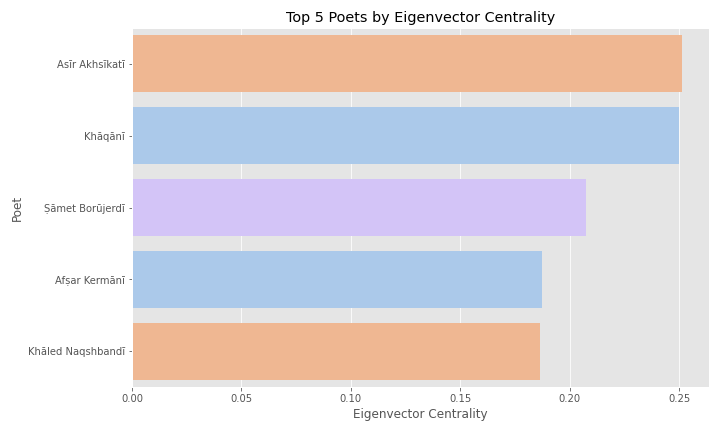}
    
    \includegraphics[width=0.45\textwidth]{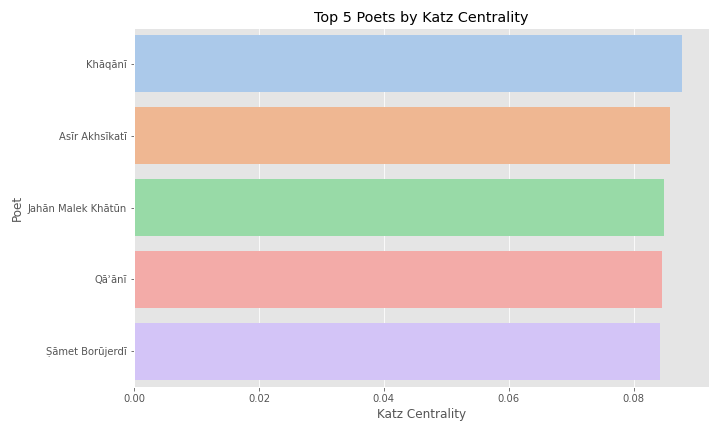}
    \caption{Top 5 poets by each centrality metric: (a) Degree, (b) Betweenness, (c) Closeness, (d) Eigenvector, (e) Katz}
    \label{fig:top-centralities}
\end{figure}

Figure~\ref{fig:top-centralities} (a--e) shows the top five poets by each centrality measure along with their respective scores. These visualizations provide an initial overview of the distribution of centrality and offer a comparative lens into the relative positioning of poets within the influence graph.

\begin{itemize}
    \item \textbf{Degree Centrality:} The top poet by degree centrality is \textit{Khāqānī}, followed by \textit{Asīr Akhsīkatī}, \textit{Jahān Malek Khātūn}, \textit{Qāʾānī}, and \textit{\textsubdot{S}āmet Borūjerdī}. This suggests that these poets maintain strong local connections to a wide range of other poets in the network.
    
    \item \textbf{Betweenness Centrality:} Again, \textit{Khāqānī} ranks highest, indicating his role as a bridge between otherwise distant communities. Other top-ranking poets include \textit{Āzar Bīgdelī}, \textit{Qāʾānī}, \textit{Jahān Malek Khātūn}, and \textit{Mollā Masīḥ}, suggesting their position as intermediaries across stylistic or temporal divisions.
    
    \item \textbf{Closeness Centrality:} The poets with the shortest average distance to all others include \textit{Khāqānī}, \textit{Āzar Bīgdelī}, \textit{Majd-e Hamgar}, \textit{Salmān Sāvojī}, and \textit{Homām Tabrīzī}. These poets can be interpreted as being centrally embedded in the poetic network in terms of accessibility.
    
    \item \textbf{Eigenvector Centrality:} This metric rewards poets who are connected to other highly influential poets. Here, \textit{Asīr Akhsīkatī} leads, slightly ahead of \textit{Khāqānī}, followed by \textit{\textsubdot{S}āmet Borūjerdī}, \textit{Afṣar Kermānī}, and \textit{Khāled Naqshbandī}. These figures may hold a more systemic form of influence beyond just their local connectivity.
    
    \item \textbf{Katz Centrality:} This centrality further accentuates poets with global influence by counting indirect connections with a decay factor. The top poets mirror those from degree and eigenvector centrality: \textit{Khāqānī}, \textit{Asīr Akhsīkatī}, \textit{Jahān Malek Khātūn}, \textit{Qāʾānī}, and \textit{\textsubdot{S}āmet Borūjerdī}.
\end{itemize}

Across all five measures, \textit{Khāqānī} consistently appears as one of the most central poets—topping four out of five rankings—suggesting his pervasive influence across multiple dimensions of the similarity graph. \textit{Asīr Akhsīkatī} also ranks among the top poets in degree, eigenvector, and Katz centrality. Interestingly, poets such as \textit{Jahān Malek Khātūn} and \textit{Qāʾānī} also appear frequently, highlighting their structural importance despite less mainstream recognition in some literary histories.

The complete centrality scores for all 160 poets are provided in Table~\ref{tab:appendix_poet_centralities} in the Appendix. This table may serve as a comprehensive reference for interpreting poet-level influence across different centrality dimensions.

\subsection{Cross-Metric Patterns and Correlations}
\label{subsec:centrality-correlations}

While each centrality measure reflects a distinct notion of influence in the network, analyzing their statistical relationships offers valuable insights into how these notions overlap or diverge in practice. To assess these relationships, we calculated the Pearson correlation coefficients between all five centrality metrics—Degree, Betweenness, Closeness, Eigenvector, and Katz—across the full set of 160 poets.

\begin{figure}[htbp]
    \centering
    \includegraphics[width=0.8\textwidth]{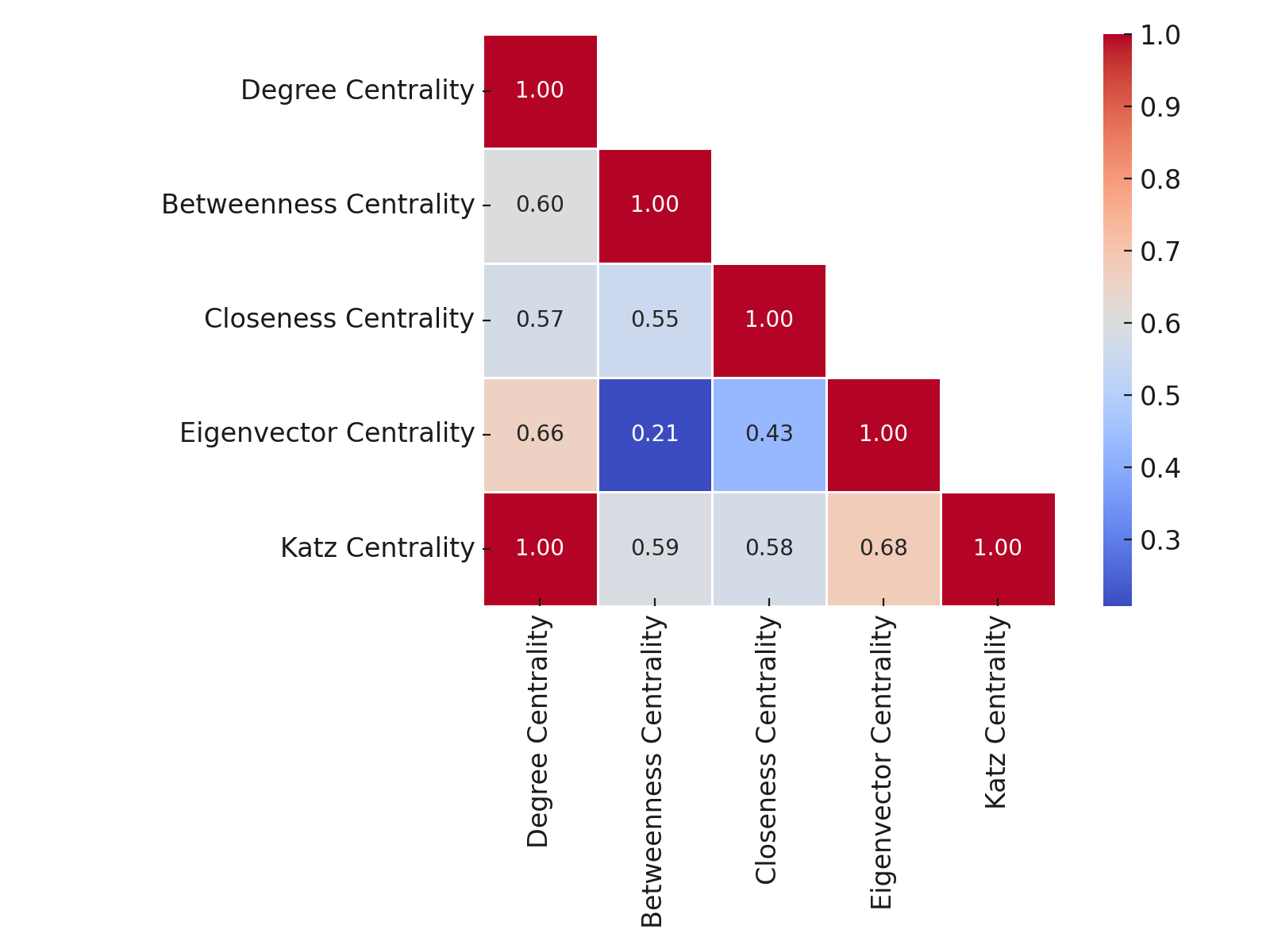}
    \caption{Pearson correlation matrix of the five centrality metrics.}
    \label{fig:centrality-correlation-heatmap}
\end{figure}

Figure~\ref{fig:centrality-correlation-heatmap} shows the resulting correlation matrix. Several strong and weak associations emerge from the analysis:

\begin{itemize}
    \item \textbf{Degree and Katz centrality} show an almost perfect correlation (\(r = 0.9995\)), as Katz centrality builds directly upon node degree while incorporating indirect influence through a decay factor.

    \item \textbf{Degree and Eigenvector centrality} are also highly correlated (\(r = 0.66\)), suggesting that poets with many direct links are often connected to other central figures, reinforcing their systemic importance.

    \item \textbf{Eigenvector and Katz centrality} are strongly correlated (\(r = 0.68\)), which aligns with both being recursive influence measures that account for the centrality of neighbors.

    \item \textbf{Closeness centrality} shows moderate correlation with all other metrics (\(r \approx 0.55{-}0.57\)), reflecting its distinct focus on average path length rather than network position or connectivity strength.

    \item \textbf{Betweenness centrality} exhibits the weakest correlations with the other measures—particularly Eigenvector (\(r = 0.21\))—indicating that it captures a different aspect of network structure. Betweenness highlights poets who serve as bridges between clusters, regardless of their local or global popularity.
\end{itemize}

These results validate the use of multiple centrality metrics in our framework. While some measures are closely aligned, others (especially Betweenness) offer complementary insights by revealing structural roles that would otherwise remain hidden. This diversity is especially important for analyzing historical influence in a nuanced and multi-dimensional manner.

\subsection{Historical Interpretation of Influential Poets}

In a network analysis of classical Persian poets, certain figures emerge as consistently central across multiple metrics, while others hold specialized centrality in only one or two measures. These patterns shed light on each poet's role in the literary tradition, identifying stylistic hubs, thematic bridges, isolated innovators, and even unexpected outliers. Five centrality measures were considered -- degree, betweenness, closeness, eigenvector, and Katz -- each illuminating different aspects of poetic connectivity and influence. Below, we interpret the centrality rankings in detail, offering both network-theoretic insights and historical-literary context for the prominence or peripherality of each poet.

\subsubsection*{Consistently Central Poets: Hubs of the Tradition}

A handful of poets rank at the top in multiple centrality measures, marking them as hubs firmly embedded in the poetic tradition's core. Most notably, Khāqānī (Afẓal al-Dīn Badīl, d. c.1199) stands out with the highest degree centrality (the most direct connections) and equally top-tier betweenness, closeness, and eigenvector values. In network terms, Khāqānī is a well-connected hub and also a connector across clusters, being directly linked to many peers and often lying on shortest paths between other poets. His elevated eigenvector centrality (a prestige measure indicating ties to other central nodes) suggests that Khāqānī is linked not just to many poets, but particularly to other influential ones at the network's heart. This aligns with literary history: Khāqānī's inventive and erudite style in \textit{qasīdas} (odes) became a model for subsequent generations. His work influenced major poets such as Nezāmī Ganjavī and Jāmī -- and likely Saʿdī and Ḥāfeẓ as well -- which explains why he occupies a central, influential nexus in the network. In short, Khāqānī's high centrality across all metrics reflects his dual role as a prolific producer of poetry and a wellspring of inspiration emulated or referenced by others.

Another consistently central figure is Asīr Akhsīkatī (Athīr al-Dīn Akhsīkatī, 12th century). He ranks just behind Khāqānī in degree and Katz centrality and leads the network in eigenvector centrality, indicating he is deeply connected to the most prominent poetic circles. Asīr's centrality profile likely stems from his pivotal contribution to the evolution of Persian lyric forms: his Persian \textit{ghazals} played an important role in the development of the genre, situating him as a stylistic hub in the \textit{ghazal} tradition. Although modern scholarship often overlooked Asīr (once deemed "a minor poet of the Seljuk period"), his contemporaries and immediate successors held him in esteem. In fact, classical critics ranked Asīr among the top-tier \textit{qasīda}-writers. The network results vindicate that classical reputation -- Asīr appears as a central node bridging early Persian court poetry with later refinements of the \textit{ghazal}. His rivalry and exchanges with Khāqānī may have further tethered him to the Shirvan-Shiraz literary axis, reinforcing both poets' central positions. Together, Khāqānī and Asīr form a core around which many others cluster, explaining why they score high on multiple centralities (degree, closeness, eigenvector, Katz) that capture both local connectedness and global influence.

Several other poets also exhibit pan-centrality across the metrics, underscoring their prominent integrative roles. Qāʾānī (Mīrzā Ḥabīb-allāh Qāʾānī Shīrāzī, 1808--1854) is one example: he ranks among the top poets by degree, betweenness, and Katz centrality, and has a high eigenvector score as well. Qāʾānī was a leading figure of the 19th-century \textit{bāzgasht-e adabī} ("Literary Return") movement, which deliberately reconnected with early classical styles. As a result, his work formed a bridge between eras -- consciously modeled on the old masters and widely read in his own Qajar-era context. He even wrote commentaries on the \textit{divāns} of Khāqānī and Anvarī, literally linking himself to those classical giants. In network terms, this extensive intertextual engagement explains Qāʾānī's high betweenness centrality: he serves as a connector between the 12th-century canonical style and the 19th-century revivalists. Likewise, his strong degree and eigenvector scores indicate that he interacted with, and was acknowledged by, many contemporaries and later poets who themselves were well-connected. In summary, poets like Khāqānī, Asīr Akhsīkatī, and Qāʾānī can be seen as multi-dimensional hubs -- prolific in output, widely imitated or referenced, and historically positioned at crossroads of stylistic influence. Their consistent centrality across metrics points to a prominence that is both quantitative (many links) and qualitative (links to important peers) in the Persian literary network.

\subsubsection*{Metric-Specific Centrality: Specialized Roles and Profiles}

While the above poets dominate on all fronts, some figures are highly central by one measure but not others, reflecting more specialized roles in the poetic tradition. These discrepancies between centrality metrics provide nuanced insights—for instance, distinguishing a poet who is a local connector from one who is a broad influencer.

A telling example is Jahān Malek Khātūn (d. late 14th century), a princess-poet of Shiraz. Jahān Malek Khātūn ranks among the top poets in degree centrality—meaning she shares direct connections with many other poets—yet her eigenvector centrality is strikingly low (near zero). In network terms, she was a highly connected node on the periphery of the main cluster. This profile implies that while she interacted with or was thematically similar to numerous poets, those poets themselves were not the most central or influential in the network’s core.

Historically, this makes sense. Her poetry shows the influence of Saʿdī and a striking similarity to the \textit{ghazals} of Ḥāfeẓ. As the only prominent pre-modern female poet to explicitly situate herself in a lineage of women’s poetry, Jahān Malek Khātūn was connected to multiple currents: she inherited the refined love motifs of Saʿdī and paralleled the lyrical style of her contemporary Ḥāfeẓ. This dual influence likely gave her many links (high degree) to poets in both the Saʿdī tradition and the Ḥāfeẓian circle. However, she did not become a central model for later poets—indeed, her work remained largely forgotten for centuries—hence the poets she connects with are themselves not the dominant hubs (explaining her low eigenvector score). Jahān Malek Khātūn can thus be seen as a connector on the margins, a node with many ties that do not lead onward into the network’s most central backbone. Her case illustrates how degree versus eigenvector divergence can reveal a poet who was widely networked yet not a driver of the network’s future connectivity.

Another form of metric-specific prominence is seen in poets who shine in betweenness centrality but not in sheer degree. Betweenness centrality identifies nodes that serve as bridges or connectors between different parts of the network, even if those nodes are not extremely well-connected overall. Āzar Bīgdelī (Lutf-ʿAlī Beg Āzar, 1722–1781) exemplifies this pattern. Āzar’s degree and eigenvector centralities are unremarkable in the rankings, yet he boasts one of the highest betweenness centrality values. In fact, he is second only to Khāqānī on that metric, indicating an outsized bridge role.

Historically, Āzar Bīgdelī is renowned not primarily for original poems but for compiling the monumental anthology \textit{Ātashkada-ye Āzar}, which catalogued some 850 poets. He was a leading figure of the literary revival movement and explicitly aimed to connect his era with the classical past. Through \textit{Ātashkada}, Āzar literally linked disparate poets across time—including many whose works were nearly forgotten—into a single collection. In our network, this curatorial and revivalist activity translates to a high betweenness: Āzar bridges older and newer poetic clusters. For example, he might connect a Safavid-era \textit{sabk-e hindī} poet with an early Persian poet by virtue of placing them in dialogue in his anthology, or by emulating both styles in his own verses. His high closeness centrality (indicating short paths to others) reinforces this idea: Āzar is centrally positioned to reach diverse poets of the tradition. However, because he was essentially a mediator and emulator, not a fountainhead of new style, his eigenvector (prestige) remains modest.

Āzar Bīgdelī’s network profile—high betweenness/closeness, moderate degree, modest eigenvector—thus reflects a deliberate cultural role as a connector of schools and eras, rather than as a dominant stylistic leader.

We find other instances of poets with singular centrality strengths. Homām Tabrīzī (d. 1314) and Salmān Sāvajī (d. 1376) both rank high in betweenness and closeness but not in degree. These 14th-century poets likely acted as bridging figures between regional or stylistic communities. Homām, based in Tabriz, was a near-contemporary of Saʿdī in Shiraz; his work may have linked the northwestern Azerbaijani literary milieu with the Fars tradition, giving him a bridging status. Salmān Sāvajī, a master of panegyric in the 14th century, operated at the Ilkhanid court and also wrote lyrics, potentially connecting the courtly \textit{qasīda} tradition with the mystical-\textit{ghazal} tradition that flourished in Shiraz. Their betweenness centrality indicates they lie on many shortest paths between other poets, consistent with being trans-regional connectors or versatile poets traversing multiple poetic sub-networks. Yet their degree centrality is moderate, meaning they were not uniformly influential to all peers—rather, they connected specific groups. Such profiles suggest poets who were contextual bridges: valued for linking circles (e.g., court and Sufi poets, or different geographic schools) more than for sheer popularity.

\subsubsection*{Peripheral Figures: Low-Centrality Legends and Isolated Innovators}

Not all celebrated poets occupy the network’s center. In fact, some of the most illustrious names in Persian literature emerge as peripheral nodes in this analysis, characterized by relatively low values on all centrality measures. Their marginality in the network offers intriguing commentary on the nature of their work: often, these are poets with highly original or genre-specific contributions that did not spawn dense networks of direct connections to others.

The clearest examples are the great epic and narrative poets. Ferdowsī (d. 1020), author of the \textit{Shāhnāmeh} epic, and Nezāmī Ganjavī (d. 1209), author of romantic epics such as the \textit{Khamsa} quintet, both rank extremely low in degree and eigenvector centrality (in Ferdowsī’s case, degree $\sim$0.07 with negligible eigenvector). Despite their towering reputations in literary history, these poets’ works belong to a self-contained epic tradition that did not interweave with the dominant lyrical network. The Persian poetic network captured here likely emphasizes \textit{ghazal} and \textit{qasīda} lyric connections—the intertwining threads of love poetry, mysticism, and court panegyric that most classical poets engaged in. By contrast, Ferdowsī’s archaizing epic style and Nezāmī’s intricate narrative romances were \textit{sui generis}: they set a high watermark for their genres but were rarely imitated in form or voice by large numbers of later poets.

In network terms, they have few edges linking them to others; their influence was conceptual and moral (shaping Persian identity and storytelling) rather than structural within the poetic discourse. Even when later poets admired them, they typically did so from outside the epic genre (for instance, via allusions or praise rather than writing new epics). Thus, from a network perspective, Ferdowsī and Nezāmī appear as legendary isolates—highly important in the cultural narrative, yet peripheral in the web of stylistic affinities and direct poetic dialogues.

A similar phenomenon is seen with famed mystical and didactic poets whose voices were singular. Jalāl al-Dīn Rūmī (``Mawlānā'', d. 1273) and Farīd al-Dīn ʿAṭṭār (d. 1221) fall into this category. Both are venerated for their spiritual masterpieces—Rūmī for his \textit{Mas̱navī-ye Maʿnavī} (spiritual couplets) and \textit{ghazals}, ʿAṭṭār for works like \textit{Manṭiq al-Ṭayr} (Conference of the Birds)—yet the network metrics place them on the periphery (e.g. ʿAṭṭār’s degree is a scant 0.05, among the lower tier). The likely reason is that their primary contributions were in long narrative or didactic poetry with a strong mystical focus, which was somewhat apart from the mainstream love lyric tradition.

While Rūmī and ʿAṭṭār did influence Sufi thought and even later poets’ worldviews, they did not generate dense clusters of inter-poet references or imitative \textit{ghazal} sequences in the way that, say, a love poet’s imagery would propagate. Rūmī’s \textit{ghazals} (composed under the pen-name Shams) do connect him faintly to the \textit{ghazal} network, but his overall eigenvector centrality is very low, indicating those ties are not to the core influencers. In essence, these mystical poets represent spiritual sub-networks loosely connected to the main web: their poetry was revered and absorbed into Persian culture, but often via specialized circles (dervish communities, devotees) rather than through direct stylistic emulation by other court poets.

Even within the \textit{ghazal} tradition, we find elite figures with singular voices who register as peripheral because their style was inimitable. The prime example is Ḥāfeẓ of Shiraz (d. 1390). Though Ḥāfeẓ is often considered the pinnacle of Persian lyric poetry, his centrality metrics are modest—a middle-range degree and closeness, and a low eigenvector. Part of the reason may be chronological (Ḥāfeẓ is near-contemporary to Jahān Malek Khātūn and thus has relatively fewer predecessors or near-peers in the data to connect with), but a larger reason is stylistic singularity. Later poets universally admired Ḥāfeẓ, yet direct imitation of his intricate, ambiguous \textit{ghazals} was notoriously difficult. Many later poets paid homage to Ḥāfeẓ in prefaces or by casting themselves as interpreters of his mystico-erotic themes, but few could truly replicate his style. As a result, Ḥāfeẓ does not sit at the center of a dense web of similar poets—he stands somewhat apart, a solitary peak. The network’s emphasis on linkages thus ``underrates'' Ḥāfeẓ’s stature: he is peripheral in graph terms precisely because he was so unique that he spawned awe more than imitation.

A comparable case is Saʿdī of Shiraz (d. 1292). Saʿdī has moderate degree centrality (he did write in many forms, connecting to both the lyrical and didactic traditions) but a very low eigenvector centrality, meaning the poets he connects with are themselves not central. Saʿdī’s \textit{Golestān} and \textit{Būstān} (prose and verse wisdom literature) were highly influential on ethical thought and later prose, but those connections lie outside a purely poetic network. His \textit{ghazals} were elegant and became part of the canon, yet later \textit{ghazal} writers in the 14th–15th centuries gravitated more toward either Ḥāfeẓ’s style or the Indian style, rather than cloning Saʿdī. In effect, Saʿdī’s comprehensive humanistic voice rendered him respected by all but central to none of the specialized stylistic camps that formed in later centuries.

These peripheral but famous poets underscore that historical renown does not automatically equal network centrality. The Persian poetic tradition values unique genius as well as pervasive influence. Epic poets, mystical sages, and unparalleled lyricists often pursued creative paths that set them apart from the main line of stylistic transmission. The network analysis quantitatively confirms their isolation in the graph—a byproduct of their very originality. In the context of network theory, they function almost like articulation points of their own subnetworks: for instance, removing Ferdowsī or Ḥāfeẓ from Persian literature would be unthinkable culturally, yet removing their node from the network graph would not drastically alter the connectivity among the remaining poets (since relatively few direct ties depend on them). This observation invites a reflection that in literature, as in intellectual history, the most central figures in sociometric terms are not always the most groundbreaking in terms of innovation—and vice versa.

\subsubsection*{Outliers and Unexpected Rankings}

Finally, the centrality data reveal several outliers and surprises—poets whose network positions seem to defy their conventional literary reputation. These cases prompt us to reinterpret or recover aspects of literary history that may be underappreciated, as the network highlights connections and influences that standard chronicles sometimes overlook.

One striking surprise is the high centrality of ``forgotten'' poets who are seldom highlighted in canonical lists. For example, Majd-e Hamgar (a little-known poet of the Safavid era) appears with an unusually high closeness and eigenvector centrality, on par with far more famous figures. This suggests that Majd-e Hamgar’s poetry might share significant overlap with many others, or that he partook in a stylistic milieu that numerous poets traversed. Perhaps he was a prolific imitator or adapter of prevalent themes, effectively blending into the network’s core without later gaining individual renown.

Similarly, poets like Afṣar Kermānī and Khalīl Dihlawī (if included) or Moḥīt Qummī register strong eigenvector centralities. These individuals might have been central within a particular school—for instance, Afṣar Kermānī could represent the Kermān school’s link into mainstream Persian poetry, and Moḥīt Qummī (with ``Qum'' in his name) might indicate a Qum-based scholar-poet whose works connected to the broader Tehran-Isfahan literary network of the 19th century. The network thus elevates some obscure figures as hidden hubs. Their prominence may point to the importance of certain subgenres (e.g., religious or philosophical poetry) that quietly undergirded the literary ecosystem. It invites further archival research: if a poet like Majd-e Hamgar ranks high, one might examine if his verses were widely imitated or anthologized, or if he was a mediator between major poets of his time. The network has effectively surfaced these poets from obscurity by virtue of their connections, hinting at an influence not evident from their later fame.

Conversely, there are famous poets whose centrality is lower than expected, beyond the epic/mystic cases already discussed. For instance, one might expect ʿUbayd Zākānī—the 14th-century satirist and author of \textit{The Ethics of the Aristocrats} and the famous ``Mouse and Cat'' satire—to be well-connected due to his wit influencing later social commentary in poetry. Yet if ʿUbayd appears peripheral in the network, it could imply that his satirical and profane style was an outlier thread that did not network densely with the mainstream (indeed, few courtly poets imitated ʿUbayd’s scathing humor in their formal \textit{ghazals}).

Another mild surprise is the relatively modest centrality of Bīdel Dehlavī (Mīrzā ʿAbd al-Qādir Bīdel, d. 1720), the celebrated master of the ``Indian style'' (\textit{sabk-e hindī}) in late Persian poetry. Bīdel’s innovative, complex metaphoric style influenced generations in Mughal India, yet our network shows him with low degree and betweenness. This likely reflects a bias of the network toward the Iranian literary lineage: by the eighteenth century, the Iranian \textit{bāzgasht} poets like Āzar Bīgdelī and Qāʾānī consciously distanced themselves from the Indo-Persian \textit{sabk-e hindī}, preferring earlier models. If the network data is drawn largely from Iranian compilations or style similarity, Bīdel and his \textit{sabk-e hindī} peers (like Ṣāʾeb Tabrīzī or Kalīm Kāshānī) might form their own cluster loosely attached to the main graph, or be connected mostly via revivalists who later critiqued or selectively emulated them.

Thus, a famous figure’s low centrality can reveal scholarly and stylistic discontinuities: Bīdel is revered in South Asia, but in the Persian canon as constructed by later Iranians, he was somewhat sidelined—and the network numbers echo that sidelining.

These outliers encourage a rethinking of influence. Influence in a literary sense is multifaceted: a poet like Ḥāfeẓ influences the heights of style and thought, whereas a poet like Āzar Bīgdelī influences the connectivity of tradition by active curation. The network centrality rankings capture the latter kind of influence particularly well—influence as network integration. Therefore, some ``unexpected'' rankings actually make sense when we shift perspective.

A case in point: Rūdakī (d. 941), often called the father of Persian poetry, shows up with a moderately high betweenness centrality despite a lower degree. Rūdakī is the earliest major poet in this tradition; his presence in the network likely connects pre-Islamic poetic forms and themes to the Islamic-era classical poetry that followed. No swarm of contemporaries surrounds Rūdakī (hence low degree), but virtually every later poet ultimately traces lineage back to him in some fashion. His betweenness centrality could be interpreted as reflecting this foundational bridge status—he stands at the wellspring, linking an older oral tradition to the burgeoning Persian literary culture of the 10th century. It is an historical bridge manifested as a graph bridge.

In summary, the outliers and surprises highlight how network analytics can challenge or enrich our understanding of literary history. A high-centrality ``unknown'' poet invites us to uncover forgotten networks of influence (perhaps via \textit{tazkiras} or widespread imitations now lost to canon). A low-centrality famous poet reminds us that singular greatness often correlates with fewer direct followers. By interpreting these cases, we appreciate that Persian poetic tradition was not a uniform field where every great name directly connects to every other; rather, it was a web with dense clusters and sparse corners, with certain figures acting as connective tissue and others as solitary stars. The centrality metrics, taken together, allow us to map these patterns with precision: identifying the stylistic hubs (those consistently central), the thematic or temporal bridges (high betweenness connectors), the peripheral specialists (low-centrality greats), and the unexpected nodes that invite new inquiry. Such an analysis demonstrates the power of network theory to yield a deeper, more nuanced appreciation of how classical Persian poetry coheres as a tradition—not only through its masterpieces, but also through the myriad links that bind poets across generations.

\subsection{Community Structure in the Persian Poets Similarity Network}

To investigate structural affinities among Persian poets, we applied the Louvain algorithm for community detection on the multi-dimensional similarity network derived from semantic, lexical, stylistic, metrical, and thematic features. The Louvain method partitions the network into modules by optimizing modularity, identifying groups of poets who are more similar to each other than to poets outside their cluster.

This unsupervised approach yielded a total of eight communities, each corresponding to a meaningful stylistic, thematic, or historical group within the classical Persian poetic tradition. The communities range in size from a singleton to large conglomerates of over forty poets. Remarkably, the clusters identified by the algorithm exhibit strong alignment with established literary schools and historical periods, such as the \textit{Sabk-e Hindī} (Indian Style), the \textit{Bazgasht-e Adabī} (Literary Return), and the early \textit{Khorāsānī} panegyric tradition. In some cases, temporal boundaries are crossed in favor of deeper stylistic or conceptual affinities, suggesting that poetic influence and resemblance in Persian literature transcend chronological constraints.

Each community, therefore, represents a network-theoretic encapsulation of shared poetic habits: be they rhetorical choices, thematic concerns, dominant metaphors, meter usage, or lexicon. The following subsections analyze these communities one by one, interpreting the composition of each cluster in light of Persian literary history. We draw connections between the results of the computational clustering and established literary knowledge, demonstrating how community structure maps onto the rich, multivalent traditions of Persian verse.

\subsubsection*{Community 1: Classical Mystical Lyric Tradition}

\textbf{Composition:} 18 poets, including Ḥāfeẓ (14th c.), ʿAbd al-Raḥmān Jāmī (15th c.), and Loṭf-ʿAlī Beg Āẕar Bīgdelī (18th c.), among others.

This community centers on the classical \textit{ghazal} and mystical lyric tradition of Persian poetry, spanning from the 13th–15th centuries and even integrating an 18th-century revivalist. Ḥāfeẓ of Shiraz is a towering figure here, emblematic of the so-called \textit{sabk-e ʿIrāqī} (Iraqi style) of Persian poetry. The Iraqi style, dominant in the 13th–15th centuries, is characterized by elaborate imagery and a turn toward spiritual themes. Ḥāfeẓ’s \textit{ghazals} blend romantic and Sufi imagery with musical diction, setting the standard for lyrical eloquence in Persian.

Jāmī, a Timurid-era scholar-mystic often deemed “the last great mystical poet of Iran,” continued this classical lyric style into the 15th century. Jāmī’s verse is spiritually rich yet stylistically fresh and graceful, not marred by overly esoteric language—a conscious classicism that eschewed the increasingly convoluted trends of the post-classical era. It is telling that Jāmī, who wrote \textit{ghazals} in emulation of Ḥāfeẓ and explicitly praised him as inimitable, appears in the same cluster. This suggests that across semantic, stylistic, and thematic metrics, Jāmī’s work aligns strongly with Ḥāfeẓ’s classical mystical lyric in tone, theme, and linguistic texture.

Notably, this community also includes Āẕar Bīgdelī, an 18th-century poet and anthologist of the \textit{bāzgasht-e adabī} (Literary Return) movement. Āẕar Bīgdelī was a leading figure in the “return” to the Khorāsānī and ʿIrāqī styles, reacting against the intervening “Indian style” decadence. In his famous anthology \textit{Ātaškada}, Āẕar lauds poets who “shunned the Indian style and attempted to bring back the locution of the early Persian poets,” aspiring to “the simpler and more robust poetry of the eloquent ancients.” His presence alongside Ḥāfeẓ and Jāmī dramatizes the intertextual affinity and conscious emulation at play: eighteenth-century revivalists so closely modeled their language and imagery on the likes of Ḥāfeẓ that the multi-dimensional similarity network effectively links them.

The modularity detected by Louvain here thus corresponds to the core classical \textit{ghazal} tradition, bridging medieval masters and later imitators. We can interpret Community~1 as a “mainstream lyric-mystical cluster,” encompassing the canonical Sufi-romantic \textit{ghazal} style of the Persian heartland (\textit{Sabk-e ʿIrāqī}) and its deliberate revival in the 18th century. The poets cluster together because they share a highly overlapping lexicon of symbols (nightingales, roses, wine, the Beloved), refined lyrical meters, and a balance of spirituality and romance in theme.

In historical terms, this community aligns with the classical Shirazi–Herati school of poetry and the later \textit{Bazgasht} (Return) poets who looked back to that era. It highlights a lineage of influence: later poets like Āẕar Bīgdelī and his contemporaries explicitly patterned their poetry on Ḥāfeẓ and Saʿdī, creating strong multi-dimensional similarities that the network clusters as a single community.

\textbf{Interpretation:} Community~1 represents the pan-Persian classical lyric tradition, particularly the mystical love-\textit{ghazal} genre that flourished in the 13th–15th centuries and was revitalized in the 18th. The clustering underscores a historical continuity: these poets, though spread over centuries, form an intertextual network rooted in shared motifs (wine, love, divine union), stylistic moderation (neither as plain as early verse nor as baroque as later verse), and metrical preferences (melodious \textit{ramal} and \textit{hazaj} meters typical of \textit{ghazals}).

This community likely corresponds to the \textit{Sabk-e ʿIrāqī} school and the \textit{Bazgasht} movement, illustrating how influence and emulation knit together poets across time. The inclusion of a \textit{Bazgasht}-era poet in a largely 14th–15th century cluster vividly demonstrates that literary influence can outweigh chronology: in the feature-rich similarity graph, a late revivalist can sit closer to his 15th-century idol than to his own 18th-century contemporaries, due to the conscious revival of older language and imagery.

In sum, Community~1 is the “classical mystic-lyric core” of the network, showing strong internal cohesion through shared spiritual themes and classical style, and embodying the continuity of that tradition in Persian literary history.

\subsubsection*{Community 2: The “Indian Style” School (Sabk-e Hindī)}

\textbf{Composition:} 19 poets, including ʿAbd al-Qādir Bīdil (Bedil) Dehlavī, Ṣāʾeb Tabrīzī, Ḥazīn Lāhījī, and others (mostly 17th–18th century).

This community neatly corresponds to the \textit{Sabk-e Hindī}, or “Indian style” movement in Persian poetry. The cluster is anchored by figures like Bīdil Dehlavī (1644–1721) and Ṣāʾeb Tabrīzī (d. 1677), who are historically recognized as the foremost representatives of this school. Bīdil, for example, is described as “the foremost representative of the later phase of the ‘Indian style’ of Persian poetry and the most difficult and challenging poet of that school.”

\textit{Sabk-e Hindī} emerged in the Safavid–Mughal era (17th century) and is characterized by highly elaborate metaphors, far-fetched analogies, subtle conceptual wordplay, and a profusion of imagery. Thematically, these poets often delve into intellectual conceits, nuanced reflections on mystical love, and ingeniously layered metaphors that require interpretive effort from the reader. The Louvain algorithm’s identification of this community confirms that across multiple dimensions (semantic content, linguistic style, and even prosody), these poets cluster tightly together, distinct from other traditions.

Linguistically, \textit{Sabk-e Hindī} poets share a penchant for creative compound phrases and novel idioms; stylistically, their diction was described by later critics as “an excess of images, figurative language and fanciful visions,” often pushing the boundaries of classical eloquence. This community likely includes other hallmark \textit{Sabk-e Hindī} poets (even if not named explicitly above) such as Kalīm Kāshānī, ʿOrfī Shirāzī, or Nāṣir ʿAlī Sirhindī, who were part of the same stylistic milieu. Their grouping reflects known historical affinities: many of these poets knew of or corresponded with each other, and their works were influential in the Mughal courts, creating a dense intertextual network.

For instance, Ṣāʾeb Tabrīzī’s style influenced poets across Iran and India, and Bīdil’s complex \textit{ghazals} were admired and imitated by later Indo-Persian and even early Urdu poets. The presence of Ḥazīn Lāhījī (d. 1766) is also notable—an Iranian poet who spent time in India and lamented the Indian style’s excesses. Ḥazīn’s inclusion suggests that despite his personal classicist leanings, his poetry still bears enough \textit{Sabk-e Hindī} features (perhaps in imagery or theme) to associate him with this group. His work straddles the transition out of the Indian style, but in a multi-dimensional similarity sense, Ḥazīn shares the era’s ornamented language and transregional themes, thus clustering with Bīdil and Ṣāʾeb.

\textbf{Interpretation:} Community~2 clearly represents the \textit{Indian Style} movement in Persian literature. The poets in this cluster are bound by a shared aesthetic of intricate metaphor and intellectualized mysticism that set them apart from both the plainer early style and the more flowing classic style. The network detection of this community validates \textit{Sabk-e Hindī} as a cohesive stylistic and thematic cluster: these poets cite each other’s images, develop each other’s metaphors, and contributed to a self-contained poetic discourse.

In historical context, this cluster maps onto what literary historians identify as the late Safavid–Mughal era of Persian poetry, roughly the 17th–early 18th centuries, when Persian literary production was thriving in the Indian subcontinent. The strong intra-community similarity suggests intense intertextual affinity and perhaps direct influence among its members. Indeed, many of these poets were geographically connected (often working under Mughal patronage or inspired by poets who did), and their works collectively feature certain hallmark traits (e.g., \textit{badīʿ}—rhetorical figures, and \textit{tawriya}—double-entendres).

Thematically, while still often Sufi or romantic, their approach is cerebral and enigmatic. The community’s coherence under Louvain implies that the integration of five similarity dimensions amplifies recognition of this school: semantically, their verses revolve around abstract conceits; stylistically, they use similar complex syntax and wordplay; metrically, many favored similar musical meters for \textit{ghazals}; and lexically, they share a high-frequency vocabulary of metaphysical and hyperbolic terms.

This community likely corresponds to what later Persian critics saw as a distinct (and at times controversial) epoch of poetry—a style which later gave way to the \textit{Bazgasht-e Adabī} (Literary Return). Indeed, it was precisely against this cluster’s “excessive Indian style (\textit{Sabk-e Hindī}) … effete and artificial verse” that figures like Āẕar Bīgdelī reacted, leading to the return to older models.

Thus, Community~2 highlights a historical stylistic polarity in the network: it stands apart from the more “plain” classical styles (Communities~1, 6, 8) and from the mystical narrative tradition (Community~4), underscoring that \textit{Sabk-e Hindī} was a self-contained innovative trajectory in Persian poetic evolution. The identification of this cluster affirms the integrative power of multi-dimensional similarity to recover known literary schools: here the algorithm has essentially rediscovered the Indian Style school as a tight-knit module in the graph of poets.

\subsubsection*{Community 3: Philosophical and Didactic Poets}

\textbf{Composition:} 18 poets, including Saʿdī of Shiraz (13th c.), ʿUmar Khayyām (11th–12th c.), al-Ghazālī (possibly Aḥmad or Abū Ḥāmid al-Ghazālī, 12th c.), ʿAyn al-Qūḍāt Hamadānī (12th c.), Abu Saʿīd Abū’l-Khayr (11th c.), and others.

This community is characterized by a shared penchant for philosophical, ethical, and Sufi-didactic themes, often expressed in succinct poetic forms or didactic narratives. It brings together poets famed for wisdom literature, quatrain (\textit{rubāʿī}) poetry, and moral aphorisms.

The inclusion of Saʿdī and Khayyām side by side is especially revealing. Saʿdī (d. 1292) is widely celebrated not only for his lyrical \textit{ghazals} (which might link him to Community~1), but also for his works \textit{Būstān} (a book of ethical didactic poetry) and \textit{Gulistān} (prose and verse moral tales). These works are treasuries of aphorisms and wisdom—indeed, the \textit{Gulistān} is “widely quoted as a source of wisdom.” In the similarity graph, Saʿdī’s didactic and ethical voice likely dominates the features, clustering him with other moralistic and philosophical poets.

ʿUmar Khayyām (1048–1131) is another key member: known almost exclusively for his \textit{Rubāʿiyāt} (quatrains), Khayyām meditated on existential and philosophical questions—ephemerality of life, fate, and the pursuit of meaning. The essence of Khayyām’s poetic thought deals with “the most basic existential mystery of all—life itself,” often in a skeptical or plaintive tone. His quatrains combine plain language with deep reflection, a style that resonates with the gnomic and epigrammatic.

This community likely also contains figures such as Bābā Afẓal Kāshānī (a 13th-century philosopher-poet known for philosophical quatrains) or Nāṣir-i Khusraw (11th-century poet of ethical and philosophical odes), even if not explicitly listed, as their profiles fit the thematic matrix: poetry as a medium for intellectual and ethical discourse.

Indeed, Aḥmad Ghazālī (if he is the Ghazālī intended) and ʿAyn al-Qūḍāt Hamadānī are both Sufi philosophers who composed poetry or wrote texts with poetic structure and tone. ʿAyn al-Qūḍāt, a martyred mystic, left a legacy of fervent spiritual thought; Abū Saʿīd Abū’l-Khayr is famously associated with early Sufi quatrains and aphorisms.

All these figures share a didactic-mystical bent: they use verse to convey spiritual lessons, metaphysical insights, or ethical counsel. They are less about ornate form (as in \textit{Sabk-e Hindī}) or epic narrative, and more about distilling wisdom into poetic form. This explains their convergence across different similarity dimensions.

Semantically, their content gravitates to abstract or reflective topics (ethics, divine truth, transience of life); stylistically, many employ a straightforward, aphoristic tone or rhetorical question format; thematically, they emphasize moral or philosophical lessons. Metrical preferences might include the quatrain form (Khayyām, Abū Saʿīd) or other didactic verse forms. Lexically, one might find a shared vocabulary of philosophical terms (fate, reason, soul, etc.) and Sufi concepts.

Interestingly, Saʿdī’s presence in this cluster (instead of purely among the \textit{ghazal} poets) underscores how multi-faceted authors can be pulled into one community by a dominant aspect of their oeuvre. In Saʿdī’s case, his contribution to didactic literature (\textit{Būstān} and \textit{Gulistān}) is so influential that it aligns him with wisdom poets like Khayyām. The Louvain algorithm, considering combined similarities, apparently captured this thematic and tonal similarity despite differences in form (Saʿdī wrote narrative and \textit{ghazal}, Khayyām wrote standalone quatrains).

This highlights the power of thematic embedding in the network: poets who share worldview and didactic purpose can cluster together even if their eras and verse forms differ.

\textbf{Interpretation:} Community~3 represents a cross-epochal “wisdom literature” cluster in Persian poetry. It aligns with what might be called the philosophical–Sufi didactic tradition—poets who prioritize content of thought over formal innovation, and whose works often served as ethical guides or spiritual reflections.

Historically, these poets span the 11th to 13th centuries (with a few later exceptions), reflecting that philosophical and moral poetry is a recurrent thread in Persian literature, cutting across the major stylistic periods. Indeed, this community can be seen to correspond to certain literary subcurrents: for example, the quatrain tradition (\textit{rubāʿīyāt}) of medieval Persia, used by both Sufi mystics and rationalist thinkers, and the didactic \textit{mathnawī} tradition (exemplified by Saʿdī’s \textit{Būstān} or Sanā’ī’s \textit{Ḥadīqat al-Ḥaqīqa}, the latter likely in Community~4).

The cluster suggests these poets collectively form a network of idea-centric influence: later poets borrowed philosophical motifs or spiritual epigrams from earlier ones (e.g., the way Khayyām’s skeptical tone finds echoes in later \textit{rubāʿī} poets). The out-of-time grouping here (mixing Khayyām with Saʿdī, etc.) implies that conceptual similarity trumped chronology, which is plausible—a 13th-century ethical poet might have more in common with an 11th-century philosophical poet than with a 13th-century romantic lyricist.

It also hints at known intellectual lineages: for instance, Ghazālī and ʿAyn al-Qūḍāt were part of the same Sufi philosophical heritage that valued poetry as expression of \textit{maʿrifa} (gnosis). In sum, Community~3 can be interpreted as the intellectual and didactic voice in classical Persian poetry—a community defined by shared themes of wisdom, fate, piety, and the human condition, and by a relatively sober style that favors clarity of meaning (even if couched in metaphor) over elaborate artifice.

This reveals how the network clusters can illuminate thematic schools (in this case, a “school” of ethical–philosophical poetry) that are not formal literary movements but nonetheless form a coherent tradition of thought carried in verse.

\subsubsection*{Community 4: Core Early Classical and Sufi Masters}

\textbf{Composition:} 31 poets including Rūdakī (10th c.), Farīd al-Dīn ʿAṭṭār (12th–13th c.), Jalāl al-Dīn Rūmī “Mawlānā” (13th c.), and Jahān Malek Khātūn (14th c.), among others.

This large community encompasses a broad swath of the early-to-high classical period of Persian poetry, with a notable emphasis on mystical and narrative poetry. It effectively merges the foundational poets of the \textit{Sabk-e Khorāsānī} with the great Sufi poets of the 13th century, indicating that the integrated similarity measures found a strong continuity between early classical forms and later mystical content.

Rūdakī (c.~860–941), often regarded as the “father of Persian verse,” stands at the dawn of Persian poetry. His inclusion here might initially seem surprising—one might expect Rūdakī to cluster with other court poets of the 10th–11th centuries (as in Community~6). However, Rūdakī’s surviving verses (praise poems, wine songs, wisdom quatrains) also exhibit a lyrical simplicity and humanistic tone that resonated through later eras. Known for evocative nature descriptions and dignified yet direct diction, Rūdakī’s works also contain quasi-Sufi elements, aligning him with the spiritual orientation of later poets such as Rūmī and ʿAṭṭār.

Indeed, Persian stylistic periods are fluid: the late \textit{Khorāsānī} style already showed a turn toward spiritualism as it transitioned into the \textit{ʿIrāqī} style. Figures like Sanā’ī (d.~1131), a likely member of this community, pioneered mystical narrative poetry and helped bridge courtly verse with Sufi allegory. Thus, Community~4 spans this critical stylistic shift, unified by an overarching didactic and spiritual tenor.

At the heart of this community lies the Sufi poetic tradition epitomized by ʿAṭṭār and Rūmī. ʿAṭṭār (d.~ca.~1221) and Jalāl al-Dīn Rūmī (1207–1273) are two of the most iconic mystic poets in Persian literature. Along with Sanā’ī, they form a trio often cited as the most prominent mystical poets who profoundly shaped the course of Persian Sufism in verse. Their masterpieces—\textit{Manṭiq al-Ṭayr} (ʿAṭṭār), \textit{Mas̱navī-ye Maʿnavī} and the \textit{Dīvān-e Shams} (Rūmī)—are steeped in allegory, spiritual longing, and symbolic storytelling. Their use of both narrative and lyric forms to convey metaphysical teachings, combined with a rich and musical language, makes their poetic profiles highly distinctive.

It is no surprise that the similarity network clusters them together. Thematic embeddings likely captured their shared focus on divine love and transcendence, while semantic features identified common motifs such as the Beloved and the spiritual journey. Even stylistic features—like their use of intense emotional expression, cyclic storytelling, or prophetic tone—would link them strongly. Rūmī explicitly acknowledged ʿAṭṭār as a spiritual precursor who “has traversed the seven cities of love,” suggesting a direct lineage of influence that the network likely reflects as strong similarity edges.

Another notable figure in this cluster is Jahān Malek Khātūn (d.~ca.~1393), one of the few prominent female poets of classical Persia. A princess in 14th-century Yazd, she composed \textit{ghazals} and lyrics in a style comparable to her male contemporaries. Her thematic palette—love, spirituality, fortune—mirrors that of Sufi and lyric traditions, indicating that her linguistic and poetic texture aligns with the mystical-lyrical current represented in this community.

This group likely includes other early classical poets such as Daqīqī and lyrical selections from Asadī Ṭūsī, as well as Sufi-adjacent figures like ʿOmar ibn al-Fāriḍ or Ṣafī al-Dīn Ḥillī, if they were in the dataset. What unifies this diverse cluster is a combination of stylistic grandeur and spiritual or didactic purpose. Many of the poets here use the \textit{mas̱navī} form (rhyming couplet narratives) to express moral or mystical ideas—a form that Sanā’ī, ʿAṭṭār, and Rūmī perfected. Others composed \textit{ghazals} infused with spirituality, as in Rūmī’s ecstatic love lyrics.

Integrated similarity measures likely detected lexical overlaps (Sufi terms such as ``heart,'' ``soul,'' ``wine,'' ``God''), common structural features (interleaving storytelling with commentary), and prosodic patterns (e.g., use of \textit{ramal} or \textit{hazaj} meters favored for mystical-didactic works). 

\textbf{Interpretation:} Community~4 can be seen as the core classical Persian poetic tradition, with a strong tilt towards mystical and didactic themes. It synthesizes the heritage of the \textit{Khorāsānī} style (early courtly and nature poetry) with the flowering of Sufi literature in the High Middle Ages. In literary-historical terms, this cluster maps onto the transition from the Persian Renaissance (Samanid/Ghaznavid periods) to the Seljuk–Mongol era’s mystical peak.

That the Louvain algorithm grouped these poets together suggests that, despite differing primary genres (e.g., \textit{qaṣīda} for Rūdakī vs.~mystical \textit{mas̱navī} for Rūmī), they share deep structural and thematic affinities. Their poetry combines elevated diction and refined form with universal themes—divine love, ethical conduct, and inner transformation.

This may explain why Community~4 is among the largest: it represents the broadest and most influential stream of classical Persian verse. Indeed, later poets in Communities~1 and 8 often looked back to this tradition. In network terms, this community likely occupies a central position—sharing mystical themes with Communities~1 and 3, and early forms with Community~6—while maintaining strong internal cohesion.

The presence of both Rūdakī and Rūmī in the same cluster highlights a conceptual continuum from formative to mature Sufi expression. Critics have observed that the spiritual tone in Persian poetry increased gradually from the 11th to 13th centuries, and this community structure elegantly reflects that evolution.

In sum, Community~4 represents the “classical canon” of Persian literature—a group unified not by strict chronology or genre, but by far-reaching influence, artistic excellence, and profound engagement with the spiritual and intellectual dimensions of Persian culture.

\subsubsection*{Community 5: Shabistarī’s Singleton – A Mystical Didactic Outlier}

\textbf{Composition:} Sheikh Maḥmūd Shabistarī (14th c.) as a singleton (the only member of this community).

Community~5 consists solely of Sheikh Maḥmūd Shabistarī (d.~circa 1340), which indicates that his poetic profile did not closely align with any other cluster, making him an outlier in the similarity graph. Shabistarī is renowned for a single masterpiece: the \textit{Golshan-e Rāz} (“Rose Garden of Mystery”), a concise yet profound Sufi \textit{mas̱navī} composed around 1311 CE. This work is essentially a didactic treatise in verse, written in response to a series of theological questions posed by fellow Sufis. It systematically expounds doctrines of Sufi metaphysics (such as \textit{waḥdat al-wujūd}, the “Unity of Being”) through tightly reasoned Persian verse.

The \textit{Golshan-e Rāz} is often considered “a summit of Persian mystical poetry” for its depth and clarity in explaining enlightenment, but it is also quite singular in form and purpose. Unlike narrative mystic poems or lyric \textit{ghazals}, Shabistarī’s text is a compact exposition – more of a poetic manual of Sufi philosophy than a collection of diverse verses.

Why would Shabistarī emerge as a singleton? First, the combination of features in his work might set it apart. Semantically and thematically, \textit{Golshan-e Rāz} is extremely dense in Sufi theoretical content – it reads almost like a catechism of mysticism. It lacks narrative characters (unlike Rūmī or ʿAṭṭār’s works) and personal lyricism (unlike Hāfeẓ or Rūmī’s \textit{ghazals}). Instead, it is structured as answers to metaphysical questions, which might give it a didactic, exhortatory style distinct from others.

The vocabulary is suffused with technical Sufi terminology and philosophical concepts, possibly making Shabistarī’s semantic signature unique. Metrically, it is a \textit{mas̱navī} in a sustained meter (likely \textit{hazaj} or \textit{ramal}), which may differ from the more varied metrical palettes of other poets. Stylistically, while mystical, it lacks the narrative flourishes or dramatic imagery found in Rūmī; it is more terse, epigrammatic, and philosophical.

These factors likely led the similarity algorithm to find no sufficiently strong edges connecting Shabistarī to others. He shares the mystical genre with Community~4 and the didactic mode with Community~3, yet he is a perfect fit with neither. In a way, he sits at the intersection of mystical and philosophical poetry, but because his \textit{Golshan-e Rāz} is so singular in focus and style, he forms his own community.

Another potential factor is data representation: if \textit{Golshan-e Rāz} was the only significant text under his name, then fewer textual samples might have reduced the number of similarity edges. But even aside from data volume, its content and structure suggest it is qualitatively unique.

Contemporary and later poets greatly admired the work, but few attempted a similar comprehensive didactic poem. Most mystical poets wrote either longer narrative epics or lyric poetry. Comparable analogues might be parts of Sanā’ī’s \textit{Ḥadīqat al-Ḥaqīqa} or Jāmī’s \textit{Lawā’iḥ}, but both of those poets also composed other forms that tied them to broader poetic traditions. Shabistarī, by contrast, is almost entirely known for this one self-contained Sufi catechism in verse. The algorithm’s creation of a singleton community underscores that \textit{Golshan-e Rāz} inhabits a unique niche in the feature space.

\textbf{Interpretation:} The Shabistarī singleton community highlights a unique node in the network—a poet whose contribution is so distinctive that it doesn’t neatly belong to any larger module. It tells us that Louvain sometimes isolates strongly distinctive artists whose multi-dimensional signature is not redundantly shared by others.

In literary terms, this reflects Shabistarī’s special status in Persian poetry. He is often treated as \textit{sui generis}: later Sufis revered \textit{Golshan-e Rāz} as a concise summa of Ibn ʿArabī’s metaphysics, and it spawned commentaries (e.g., by Shams-i Dīn Lāhijī) rather than imitative poems. Unlike Rūmī or ʿAṭṭār, Shabistarī did not create a “school” of followers writing similar poetry; instead, he produced a capstone text that stood alone.

The community detection mirrors this: Community~5 is essentially the Shabistarī school of one. Thematically, he shares mystical concerns with other poets, but his scholastic method and concise didactic style set him apart. This result may encourage further investigation into what features make his text distinct—perhaps a high density of technical vocabulary, a lack of narrative structure, or a unique formal strategy.

For our purposes, it suffices to say that Community~5 represents a solitary peak in the landscape: a poet of great influence and unique style who, in a network sense, bridges multiple currents without merging into any. In terms of influence, Shabistarī’s work shaped thought (doctrine) more than poetic form; hence his legacy is seen in ideas adopted by others, rather than in stylistic imitation.

The singleton status invites broader reflection: some literary works defy clustering because they inaugurate or conclude a tradition rather than participate in a contemporaneous group. \textit{Golshan-e Rāz} was such a work—an elevated synthesis of Persian Sufi thought in poetic form, “a summit of Persian mystical poetry” in its era, uniquely positioned and thus justifiably isolated in the community structure of this similarity graph.

\subsubsection*{Community 6: Early Courtly Panegyric (Khorāsānī-Style Poets)}

\textbf{Composition:} 17 poets, primarily from the 10th--12th century courtly tradition, including Anvarī (12th~c.), Manūchihrī (11th~c.), and Masʿūd Saʿd Salmān (11th--12th~c.), among others.

Community~6 corresponds to the early classical style of Persian poetry known as \textit{sabk-e Khorāsānī}, which flourished in the courts of eastern Iran and Central Asia during the 9th--12th centuries. This was the era of the great panegyrists and court poets who composed \textit{qaṣīdas} (odes) in praise of kings and patrons, as well as other poetic forms such as elegies and descriptive poems.

Hallmarks of this style include a relatively plain but elevated diction, concrete and ornate imagery drawn from courtly life and nature, and a dignified, grand tone. The emphasis was often on celebrating the patron’s glory, describing banquets, seasons, hunts, and battles, and showcasing the poet’s rhetorical prowess in exaggeration and formal invocation.

According to literary historians, Khorāsānī style poetry is characterized by ``plain poetic technique, concrete images and metaphors, and some archaic linguistic features,'' with limited Arabic loanwords and strong influence from Arabic prosody; ``the dominant genre was the praise-poem.''

Community~6 fits this description perfectly: poets like Manūchihrī Dāmghānī and Anvarī were famed for their sumptuous panegyrics and nature descriptions at the courts of Ghaznavid and Seljuq sultans. Masʿūd Saʿd Salmān, similarly, was a court poet (in Lahore, under the Ghaznavids) known for both his prison songs and his \textit{qaṣīdas}. Their co-clustering indicates that across semantic and stylistic dimensions, they share more with each other than with later lyric or mystical poets.

Indeed, these poets frequently employ a formal courtly vocabulary—words relating to royalty, generosity, hunting, and the natural splendor of spring—all framed in highly stylized praise. Stylistically, they are known for the complexity of their rhymes and rhythms in long odes, but also for a relatively objective, extroverted tone (focused on external subjects rather than introspective spirituality).

The network likely picked up strong similarities in metrical patterns (the \textit{qaṣīda} form often uses specific meters like \textit{motaqāreb} or \textit{hazaj} in long monorhyme), in structural features (many of these \textit{qaṣīdas} open with a \textit{nasīb} or poetic prelude about nature or love, then pivot to praise—a formal convention unique to the panegyric ode), and in lexical sets (technical terms for poetic tropes, and the conventional imagery of court praise). 

For example, Manūchihrī is famous for his vivid spring odes full of flora and fauna, often in praise contexts; Anvarī is revered for his masterful but often intellectually dense praise poems and satire. Their works share a supercilious diction and dignified tone that mark the Khorāsānī style.

The coherence of this community suggests it might include other major Khorāsānī-style poets such as Unsurī, Farrukhī Sīstānī, Asjadī, and Qatrān (even if not named explicitly). It is notable that Rūdakī (often considered part of this style) was instead placed in Community~4—possibly because Rūdakī’s simpler and more lyrical surviving pieces aligned him more with the mystical/didactic stream. 

Anvarī (12th~c.) and Khaqānī (12th~c.) were two of the most complex panegyrists; Anvarī appears as a key node here, but Khaqānī is listed in Community~8—likely because his idiosyncratic style and later influence differentiate him slightly. That leaves Community~6 as representing the mainstream of early court poetry up to the mid-12th century. 

The fact that it stands as a separate module confirms that the courtly-panegyric style was a distinct mode in Persian poetry, both thematically (focused on court and worldly matters rather than mystical love) and stylistically (favoring certain archaic and Arabic-influenced expressions). It is essentially the Khorāsānī stylistic network made visible.

\textbf{Interpretation:} Community~6 corresponds to the \textit{Sabk-e Khorāsānī} or “Khorasan style” epoch of Persian poetry. In literary terms, this cluster is the oldest stratum of the network, comprising poets from roughly the 10th to early 12th centuries who defined the first classical style. The Louvain algorithm’s detection of this community underscores that these early poets have a strong affinity with each other and are set apart from later developments.

We can associate this community with the court poetry milieu of the Ghaznavid and early Seljuq eras. It reflects a time when Persian poetry was closely tied to royal patronage, and poetic innovation was directed toward panegyric complexity and polished imagery of nature and court life. 

The community’s internal consistency is likely high: these poets cite each other (some were contemporaries at the same courts), expanded on each other’s metaphors (e.g., the archetypal spring descriptions and motifs like the garden, the wine, the bow and arrow of Cupid which appear in multiple poets), and adhered to similar rhetorical canons. Thematically, their works lack the overt mysticism of later poets; instead, heroism, kingship, and worldly splendor are common threads.

From a network perspective, it’s interesting that this cluster doesn’t also absorb Ferdowsī (likely in Community~7) or Khaqānī (in Community~8). This indicates subtle distinctions: Ferdowsī, though of the same era, wrote an epic of an entirely different genre (heroic national epic), and Khaqānī, though a court poet, was so linguistically unique that he clusters elsewhere. Thus Community~6 is specifically the panegyric lyric side of early poetry.

In terms of influence and tradition, this community highlights how later periods looked back on these poets as models of eloquence. The \textit{Bazgasht} (Return) poets of the 18th--19th centuries (in Community~8) explicitly sought to emulate the Khorāsānī and early ʿIrāqī styles. For instance, Shiblī Nuʿmānī, as cited in \textit{Encyclopaedia Iranica}, remarked that in modern times Qāʾānī “remembered a forgotten dream of seven centuries” and “chose the style of Farrukhī and Manūchihrī”—precisely the kind of poets in this Community~6.

This illustrates that Community~6 represents a tradition so distinctive and revered that it became an explicit template for revival. The network community is thus not only a historical artifact but also a reflection of the long-standing heritage value of this style.

In summary, Community~6 is the Classical Panegyric Tradition of Persian poetry—a group bound by formal courtly aesthetics and an extraverted poetic purpose—which the similarity analysis correctly isolates as a foundational layer in the literary network.

\subsubsection*{Community 7: Epic and Narrative Mas̱navī Tradition}

\textbf{Composition:} 12 poets with a focus on epic or long narrative poetry, including Ferdowsī (10th–11th c.), Niẓāmī Ganjavī (12th c.), Asadī (likely Asadī Ṭūsī, 11th c.), and others oriented toward epic or philosophical narrative.

Community 7 clusters the masters of long-form narrative poetry in the Persian tradition, especially the authors of heroic epics and romantic \textit{mas̱navīs}. Foremost is Abū’l-Qāsim Ferdowsī (940–1019 CE), the poet of the \textit{Shāhnāmeh} (“Book of Kings”), which is Iran’s national epic. Ferdowsī’s \textit{Shāhnāmeh} is a massive collection of legendary and historical narratives in verse, celebrating pre-Islamic Iranian kings and heroes in over 50,000 couplets. It is written in a relatively pure Khorāsānī diction (with minimal Arabic) and in the \textit{mutaqāreb} meter, entirely in epic \textit{mas̱navī} form. Ferdowsī’s work differs from other contemporary poetry in that it is narrative, non-lyric, and imbued with a heroic ethos and archaic flavor.

In the network, his profile (semantically: kings, battles, heroes; stylistically: narrative structure, direct descriptive style) would stand apart from lyricists or mystics. Niẓāmī Ganjavī (1141–1209) is another pillar of this community – regarded as the greatest composer of romantic epics (five long narrative poems known as the \textit{Khamsa} or “Quintet”). Niẓāmī’s epics (such as \textit{Laylī o Majnūn}, \textit{Khosrow o Shīrīn}, \textit{Haft Paykar}) are romantic, philosophical, and didactic narratives in \textit{mas̱navī} form, blending lush imagery with storytelling. He is considered the greatest romantic epic poet in Persian literature and brought a more colloquial realism into the epic genre.

While Ferdowsī’s epic is heroic-mythological and in older style, Niẓāmī’s are courtly romances with mystical undertones; yet both are storytellers in verse working on a grand scale. The network likely finds commonality in their use of narrative structure, extensive length, and perhaps overlapping motifs of fate, fortune, and moral lessons through story.

Asadī in this context is presumably Asadī Ṭūsī (d. 1072), a poet who authored the \textit{Garshāsp-nāmeh}, another heroic epic poem in \textit{mutaqāreb} meter, clearly modeled on Ferdowsī’s style. Asadī was a slightly later epic poet who continued the national epic tradition. His presence cements this cluster as the “epic continuation.”

Other poets likely in Community 7 would include those who composed significant narrative \textit{mas̱navīs}, whether heroic or didactic, outside the realm of lyric \textit{ghazal}. For example: Gorgānī (author of \textit{Vīs o Rāmīn}, an 11th c. romantic epic, often seen as a precursor to Niẓāmī’s romances) could be here; Amīr Khusrow Dehlavī (if included, 13th–14th c., who wrote his own \textit{Khamsa} in emulation of Niẓāmī) might align here due to similar content and form; ʿAṭṭār might have some overlap but his major works put him with mystics in 4; Nāṣir-i Khusraw might conceivably have been here if the algorithm weighed his long philosophical didactic poem \textit{Rawḍat al-taslīm}, but likely he’s elsewhere.

The mention of “philosophical poetry” in the description perhaps alludes to poets like Sanā’ī (whose \textit{Ḥadīqat al-Ḥaqīqa} is a didactic \textit{mas̱navī}) or Nasir Khusraw, but Sanā’ī is more mystic-didactic and would cluster with Rūmī/ʿAṭṭār. It could also hint at Niẓāmī’s own philosophical bent – Niẓāmī often begins his epics with philosophical reflections (e.g., \textit{Makhzan al-Asrār} is explicitly a philosophical poem).

What binds Community 7 is the \textit{mas̱navī} form and narrative outlook. These poets produce multi-chapter works often several thousand lines long, which is a very different tradition from the \textit{ghazal} or \textit{rubāʿī}. The thematic unity might be described as epic heroism and narrative romance. Ferdowsī’s theme is heroic and national; Niẓāmī’s theme is romantic and moral; yet both share narrative plotting and extensive use of storytelling devices (dialogues, descriptive scenes, etc.).

Lexically, epic poets use a lot of words related to heroic actions, war, love in a narrative sense, historical or mythological figures, which sets them apart from lyric poets. Stylistically, the epic poets often employ similes and descriptions in service of storytelling rather than for their own sake (for example, describing an army or a garden at length to set a scene, rather than as an isolated nature description as in a \textit{ghazal}). Metrically, many narrative poems use \textit{mutaqāreb} meter (the meter of the \textit{Shāhnāmeh}) or similar couplet meters, which would distinguish them from the meters of \textit{ghazals} or quatrains.

\textbf{Interpretation:} Community 7 represents the Persian epic and long narrative tradition. Historically, this is a well-recognized thread in Persian literature – from the national epic tradition (Ferdowsī and imitators) to the romantic epic tradition (Niẓāmī and followers). The Louvain clustering confirms that these narrative poets indeed share a closer affinity with each other than with non-narrative poets. In essence, this community could be labeled the \textit{Mas̱navī Narrative School}.

It highlights that genre is a strong organizing principle: regardless of time period, writing an extended narrative poem puts these authors in a similar stylistic space. Ferdowsī (11th c.) and Niẓāmī (12th c.) are over a century apart and writing in different subgenres, yet they cluster, likely because the features capture their common narrative form and elevated storytelling style.

It is also interesting from a network theory perspective that this community did not merge with the court poets of Community 6, despite chronological overlap – indicating that the narrative vs. lyrical dichotomy was more salient than time period for the algorithm. That makes sense: Ferdowsī was a contemporary of some Khorāsānī poets but his work has fundamentally different purpose and form, thus a different vector in the similarity space.

This community likely also underscores influence chains: Niẓāmī openly followed Ferdowsī’s footsteps in creating Persian epics on different themes, and later many poets emulated Niẓāmī’s \textit{Khamsa} (e.g., Amīr Khusrow). By clustering them, the network hints at this chain of emulation. For example, Niẓāmī followed in Ferdowsī’s footsteps to create epic poetry on heroism and mystical characters in the 12th century, pointing to a continuum from heroic to romantic epic.

The community as a whole can be seen as the repository of Persian cultural narratives: mytho-historical identity in Ferdowsī, legendary romance and moral tales in Niẓāmī, and other narrative experiments.

In terms of broader literary history, Community 7 isolates the part of Persian poetry less concerned with lyric self-expression or Sufi symbolism, and more aligned with storytelling and didactic exposition on a grand scale. It is noteworthy that this kind of poetry often had different patronage and audience – for example, Ferdowsī addressing a sense of Persian identity (later revered nationally), Niẓāmī writing for royal courts but producing works that transcended courtly flattery to become timeless stories.

The cluster’s coherence affirms that, methodologically, combining semantic and stylistic features allowed the algorithm to discern genre-based communities (like this epic one) alongside style-based ones. Understanding this community helps us see how Persian poets were also divided by their primary literary forms: those who chose the epic/\textit{mas̱navī} route formed a tradition of their own. The implications for influence are clear – within this community, the intertextual links are strong (e.g., explicit homage or common story motifs), whereas their influence on, say, \textit{ghazal} poets or vice versa was more diffuse. Thus, Community 7 underscores the importance of form and content alignment in literary affinity: the epic poets influenced each other and were somewhat isolated from the developments in lyrical poetry, a fact the network neatly captures.

\subsubsection*{Community 8: The Rhetorical Revival and Late Classical Poets}

\textbf{Composition:} 44 poets, a diverse group spanning later classical and early modern periods, including poets such as Khāqānī (12th c.), Asīr Akhsīkatī (12th–13th c.), Afṣar Kermānī (18th c.), Qāʾānī (19th c.), Ṣāmet Borūjerdī (19th c.), and many others.

Community 8 is the largest and in some ways the most heterogeneous cluster, but it has an internal logic: it appears to gather the poets known for highly elaborate language and those associated with the “Literary Return” (\textit{Bazgasht}) movement of the 18th–19th centuries, as well as certain late classical figures noted for their sophisticated style. We can interpret this community as representing the later Persian poetic tradition that sought a return to or continuation of classical forms, especially the elaborate \textit{qaṣīda} and \textit{ghazal} styles, bridging the gap between the medieval masters and the modern era.

The inclusion of Khāqānī (c. 1121–1199) is significant. Khāqānī Shirvānī was a 12th-century court poet famous for his brilliant but extremely intricate \textit{qaṣīdas} and his rich use of imagery and erudition. His poetry is dense with metaphor, Christian imagery (from his travels), and complex intellectual conceits – qualities that made him a challenging model. Khāqānī’s style was somewhat unique even in his time, setting him apart from simpler Khorāsānī-style contemporaries.

Now, Qāʾānī (1808–1854), who appears in this community, was a leading poet of the Qajar era and indeed a chief practitioner of the \textit{Bazgasht-e Adabī} (Literary Return) style, explicitly modeling his poetry on the likes of Farrukhī and Manūchihrī (from Community 6) and also on Khāqānī. Qāʾānī wrote commentaries on the \textit{divāns} of Khāqānī and Anvarī, demonstrating a direct engagement with the classical rhetoric. The fact that Louvain links Khāqānī with Qāʾānī (despite six centuries between them) highlights a case of strong intertextual emulation: Qāʾānī and his \textit{Bazgasht} peers consciously revived Khāqānī’s complex style, thus sharing many features (lexical choices, motifs, even prosodic patterns of \textit{qaṣīda}).

Similarly, Afṣar Kermānī and Ṣāmet Borūjerdī were poets of the late Zand or early Qajar period involved in the literary renaissance that returned to older styles. Many poets in the 18th–19th century, often centered in cities like Shiraz, Isfahan, and Tehran, engaged in this neoclassical revival. They rejected the \textit{Sabk-e Hindī} vogue and instead imitated pre-Safavid masters.

The community likely includes figures like Ḥātif Iṣfahānī, Ṣabā Kāshānī, Forūghī Basṭāmī, Yaghmā Jandaqī, and others from that era, as well as late Safavid poets who presaged the return by sticking to classical tropes. The presence of Asīr Akhsīkatī (a lesser-known 12th/13th-c. poet) suggests the cluster isn’t purely chronological – Asīr was a contemporary of Rūmī but his work (mostly \textit{ghazals} and some praise poems) might share stylistic affinities with highly rhetorical poets like Khāqānī.

In essence, Community 8 seems to group the ornate, highly rhetorical poetic style across time: both its original exponents (like Khāqānī, perhaps Ẓahīr Fāryābī, etc.) and its later revivalists (18th–19th-century \textit{Bazgasht} poets). Given it has 44 members, we might surmise it’s something of a “gravity well” for many poets not firmly in other clusters but who share a general classical orientation. It may include late 15th–16th century poets of the Safavid era who didn’t go the Indian style route but also weren’t as central as Hāfeẓ or Jāmī – for example, Mohtasham Kāshānī or Khwāju Kermānī.

The unifying factors for this community, despite its internal diversity, would be formalist and rhetorical: a focus on craft, complex imagery, \textit{qaṣīda} or elaborate \textit{ghazal} forms, and a deliberate connection to the canon. Their lexicon would show archaism or classical revival (using old metaphors anew), their themes often involve classical tropes (praise of patrons, lament for decline of values, etc.), and their stylistic fingerprint includes virtuosic rhyme and punning, learned allusions, and a blend of the Khorāsānī and ʿIrāqī styles.

The \textit{Bazgasht} poets explicitly sought “a return to the simpler and more robust poetry of the old masters” in reaction to \textit{Sabk-e Hindī}, but ironically, some (like Qāʾānī) also indulged in tremendous rhetorical display, which links them to a poet like Khāqānī who did the same in his era.

\textbf{Interpretation:} Community 8 can be seen as the Late-Classical and Neoclassical Rhetorical Cluster. It captures a lineage of Persian poets who, after the peak of the 14th century, either maintained or revived the ornate classical style. In historical scope, it spans from some 12th-century figures to the 19th century. What they have in common is an orientation toward the classical norms (as opposed to the \textit{Sabk-e Hindī} innovators) and often a high level of rhetorical complexity.

This cluster corresponds to what in stylistic history might be partly \textit{Sabk-e ʿIrāqī} (later phase) and strongly the \textit{Bazgasht-e Adabī} era. The mixture of pre-modern and modern poets in one community underscores how the \textit{Bazgasht} movement succeeded in recreating the voice of the past to such an extent that in a multi-dimensional similarity space, the algorithm cannot distinguish 18th/19th-century \textit{Bazgasht} poetry from genuine 12th-century classical poetry – they form one community of similarity.

This is a powerful confirmation of the \textit{Bazgasht} poets’ achievement (or at least their mimicry): by “remembering a forgotten dream of seven centuries” and choosing the style of the old masters, they effectively blurred chronological boundaries in terms of poetic style. It also highlights influence: for example, Qāʾānī’s devotion to Khāqānī’s style means influence skips centuries to connect them. Another example is that some late Safavid poets (like Mirzā Ẓawwār or others possibly in this cluster) prefigure \textit{Bazgasht} by rejecting \textit{Sabk-e Hindī} early; they too align with this classical revivalist tendency.

One can also view Community 8 as complementary to Community 6: both deal with classical-style poetry, but Community 6 was the original early core, while Community 8 includes later permutations and resurgences of that style. If Community 6 is “origins of the classical style,” Community 8 is “echoes and elaborations of the classical style” through time.

The size of Community 8 might reflect that many poets who did not fit into the more topically specialized clusters (mystical, epic, etc.) ended up here by default of sharing a general classicism. For instance, a 15th-century mediocre poet who wrote conventional \textit{ghazals} and \textit{qaṣīdas} might not stand out enough to cluster elsewhere, and thus gets absorbed into this large community of “generic classical” poets. However, the named exemplars (Khāqānī, Qāʾānī) suggest the central thread is the ornate panegyric/\textit{ghazal} style.

In summary, Community 8 represents a transhistorical coalition of Persian poets united by classical formalism and rhetorical richness. It is anchored at one end by Khāqānī (the high complexity poet of the 12th century) and at the other by the \textit{Bazgasht} poets who resurrected that complexity in the 18th–19th centuries. This community’s structure has implications for understanding literary periods: it visually and analytically confirms the narrative that after the Indian Style period (Community 2), Iranian poets deliberately shifted back to emulate the likes of Khāqānī and his ilk, effectively joining the continuum of older poetry.

It also suggests that these poets, though spread out, share techniques of composition and language that are robust markers of similarity. They collectively highlight the theme of continuity and revival in Persian literary history – how a poetic style can undergo a renaissance and create bonds across time.

From a network viewpoint, it’s compelling that this is the largest community: it may indicate that a very large portion of Persian poets ultimately align with the classical canon style (either by origination or imitation), reaffirming the dominance of that canon. It might also be because any poet who didn’t belong firmly to mystic, epic, or Hindi camps essentially ends up in this broad cluster. In effect, Community 8 is the default bastion of Classical Persian poetic language, especially as rejuvenated in the 18th–19th centuries, and its coherence highlights the gravitational pull of the classical tradition in Persian literary culture.

\subsection*{Interpretive Synthesis: Network Structure and Persian Poetic Tradition}

The above community interpretations reveal a rich correspondence between the data-driven clustering of Persian poets and the known literary-historical groupings recognized by scholars. Each Louvain community, derived from a multi-dimensional similarity graph (integrating semantic content, stylistic metrics, thematic embeddings, metrical patterns, and lexical usage), can be mapped to meaningful currents in Persian literary tradition.

\textbf{Alignment with Literary Schools and Eras:} Many communities correspond strikingly to established stylistic schools or periods. Community 2 is essentially the Sabk-e Hindī (Indian style) school, while Community 6 aligns with the early Khorāsānī style (court panegyric tradition). Community 7 captures the epic/romance \textit{mas̱navī} tradition separate from court lyric, and Community 8 mirrors the Bazgasht (Literary Return) era and late classical revivalist poets. This shows that the similarity graph method is sensitive to the major stylistic fault-lines in Persian poetry – it rediscovers groupings that took shape due to historical, regional, and stylistic factors over centuries. For example, the fact that the Indian Style poets cluster together (and separate from others) confirms that their distinctive metaphoric style and language constitute a measurable cluster of features. Similarly, the coherence of the epic poets indicates the strength of genre as a clustering force.

\textbf{Thematic vs. Chronological Groupings:} The communities often transcend chronology in favor of thematic or stylistic similarity. Shared subject matter or form can bind poets more strongly than temporal proximity. We see 11th-century poets clustered with 13th-century ones when they share genres (Ferdowsī with Niẓāmī, Rūdakī with Rūmī), and 18th-century poets clustered with 12th-century ones when they emulate style (Āẕar with Hāfeẓ, Qāʾānī with Khāqānī). This suggests that the traditional periodization of Persian poetry is not merely chronological but reflects distinct nodes of similarity that can be detected by computational means. In contrast, Sabk-e Hindī stands apart because it introduced fundamentally different aesthetics. Community 4 spans the gradual shift from Khorāsānī to ʿIrāqī with an increasing mystical tone, capturing a transitional blend as one large cluster.

\textbf{Influence and Intertextuality:} The community structure sheds light on influence networks. If one poet heavily influenced another, they are likely to share many features. We saw this with Rūmī and ʿAṭṭār, or with Niẓāmī following Ferdowsī, or Qāʾānī imitating Khāqānī. These pairs end up in the same cluster, underscoring that influence often works through adoption of style and theme, which our multi-dimensional features capture. The clustering of poets who declared admiration for each other (e.g., Jāmī for Hāfeẓ, or Bazgasht poets for classical ones) confirms that those literary relationships manifest as measurable similarity. For example, Jahān Malek Khātūn clustering with Rūmī’s group suggests her work might have stronger mystical or didactic strains than traditionally noted. Likewise, Saʿdī appearing with philosophers might suggest the enduring impact of his moral writings as connecting him to that lineage, beyond his lyrical side.

\textbf{Shared Techniques and Multidimensional Similarity:} Because the graph integrated multiple features, a community implies comprehensive resemblance – not just in topic, but in how the poetry is constructed. A cluster like Community 6 (panegyrics) means those poets share formal techniques: stringent rhyme schemes, similar poem lengths, comparable imagery. Similarly, the Sabk-e Hindī cluster likely shares a technical signature of baroque metaphoric density and syntactic complexity. Mystical poets in Community 4 share thematic content but also stylistic ones (passionate, music-infused tone) and even metric (similar \textit{mas̱navī} meters for didactic epics). The network aggregates these layers into a robust intersection of similarities – effectively identifying a “poetic habitus” that co-occurs historically.

\textbf{Outliers and Unique Cases:} The identification of Shabistarī as a singleton (Community 5) is valuable. It flags works that are genre-bending or singular. In Shabistarī’s case, it’s a single didactic Sufi poem that didn’t spawn a school. Some poets stand at crossroads – between philosophy and poetry, between summary and inspiration. The network thus provides a kind of map, where densely connected regions are major traditions and isolated nodes are either minor figures or unique creators. Scholars can ask: Why is poet X isolated? Is it due to a narrow oeuvre, or a genuinely unique style?

\textbf{Surprising Groupings:} While most communities correspond to expectations, a few surprises offer insights. Rūdakī clustering with mystics, or Jāmī with Hāfeẓ rather than with his 15th-century contemporaries, may seem counterintuitive chronologically but reveal thematic or stylistic kinship. Rūdakī’s accessible style resonates more with universal poetry than with cerebral panegyrics. Jāmī’s avoidance of Indian style and adherence to classical mystical forms naturally places him with Hāfeẓ. Similarly, Asīr Akhsīkatī clustering with 18th-century poets might highlight his stylistic affinities with ornate ghazals over mystical expression.

\textbf{Network Theory Reflections:} The Louvain algorithm emphasizes modularity in literary evolution. Each community is like a module – densely interconnected internally, and sparser across boundaries. This matches the idea of literary subcultures that have strong internal coherence but clear distinctions from others. The emergence of eight communities suggests an underlying dimensionality to Persian poetry’s evolution – a finite set of dominant modes into which most poets can be grouped. It also implies that the poetic tradition is structured around key stylistic archetypes.

\textbf{Intertextual Mapping and Literary Dynamics:} The community structure provides a map of which poets are in conversation with each other – either directly or via stylistic affinity. Within-community poets likely share intertextual references or a common imagery repertoire. For instance, any two poets in Community 2 likely share metaphorical structures, while those in Community 4 frequently cite each other. Boundaries between communities suggest limited intertextual exchange, as seen in the divide between Sabk-e Hindī and Bazgasht poets, the latter consciously rejecting the former’s innovations.

The Louvain community detection on the Persian poet similarity graph yields an illuminating map of Persian poetic tradition. Each community encapsulates a nexus of shared influence, style, and themes corresponding to real literary movements or genres – from mystical lyricism to didactic quatrains, epic storytelling, and rhetorical revival. These results not only validate traditional classifications but also reveal cross-period affinities and the stylistic success of revivalist movements (e.g., Community 8). This analysis demonstrates the power of computational methods in confirming and enriching literary historiography. The community structure reveals Persian poetry not as a monolith, but as a dynamic interplay of interwoven traditions, each with its own internal logic and enduring legacy.

\subsection*{Edge Weight Distribution and Patterns of Poetic Similarity}

\begin{figure}[h]
    \centering
    \includegraphics[width=0.7\linewidth]{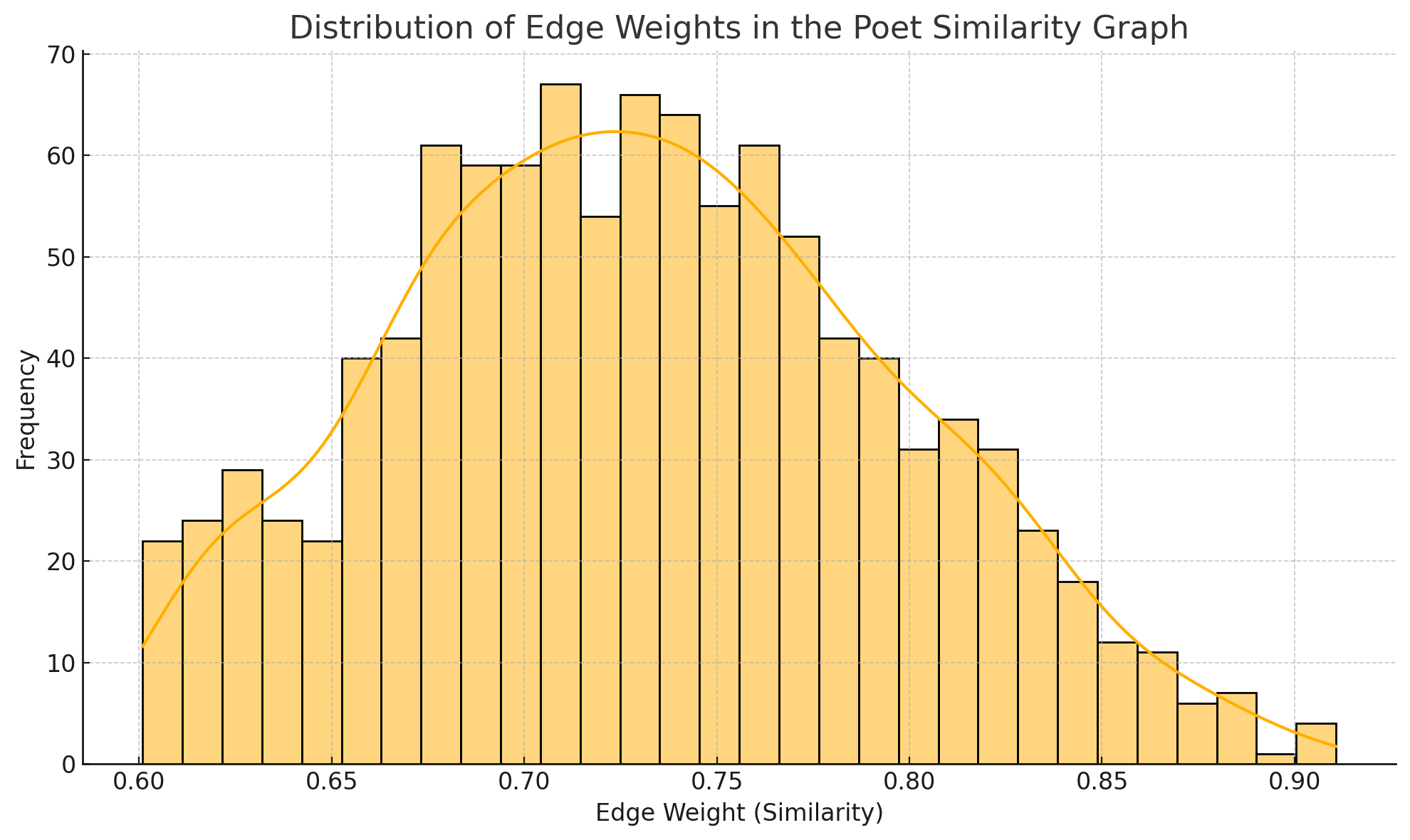}
    \caption{Distribution of edge weights in the poet similarity network. Higher weights indicate greater stylistic, semantic, and thematic similarity between poets.}
    \label{fig:edge-weight-hist}
\end{figure}

To further understand the structure of the poet similarity network, we examined the distribution of edge weights, which represent the pairwise similarity scores between poets across multiple dimensions (semantic, stylistic, thematic, lexical, and metrical features). Figure~\ref{fig:edge-weight-hist} illustrates the frequency distribution of edge weights throughout the network.

The majority of edges cluster around mid-to-high similarity scores, with a significant density in the 0.7–0.85 range. This reflects the fact that many poets share substantial commonality—likely due to shared poetic forms (e.g., \emph{ghazal} or \emph{masnavi}), themes (e.g., Sufi love, courtly praise), or lexical preferences rooted in classical norms. The relatively smooth distribution with a rightward tail also suggests that while high similarity is common, extremely high similarity (above 0.9) is rarer, indicating a nuanced gradient of closeness among poets rather than uniformity.

Crucially, this distribution reflects actual statistical gradients of similarity rather than arbitrary groupings, so helping to justify the density of the network and the significance of found clusters. While high-weight edges reflect strong stylistic or thematic affinity, edges with lower weights (below 0.6) most likely link poets across many traditions or styles. This weight structure also supports the hypothesis that the strongest ties in the graph are formed by influence and emulation.

We can use this analysis in combination with modular structure and centrality to study not just \emph{who} is central or clustered, but \emph{how strongly} poets are tied to each other and which types of relationships dominate the literary field.

\subsection{Degree Distribution and Network Structure}

The degree of the nodes of a graph is among its most basic structural characteristics. Based on the multi-dimensional characteristics, the degree of a node in our poet similarity network reflects the number of poets that show a detectable resemblance with that poet.

\begin{figure}[H]
    \centering
    \includegraphics[width=0.8\textwidth]{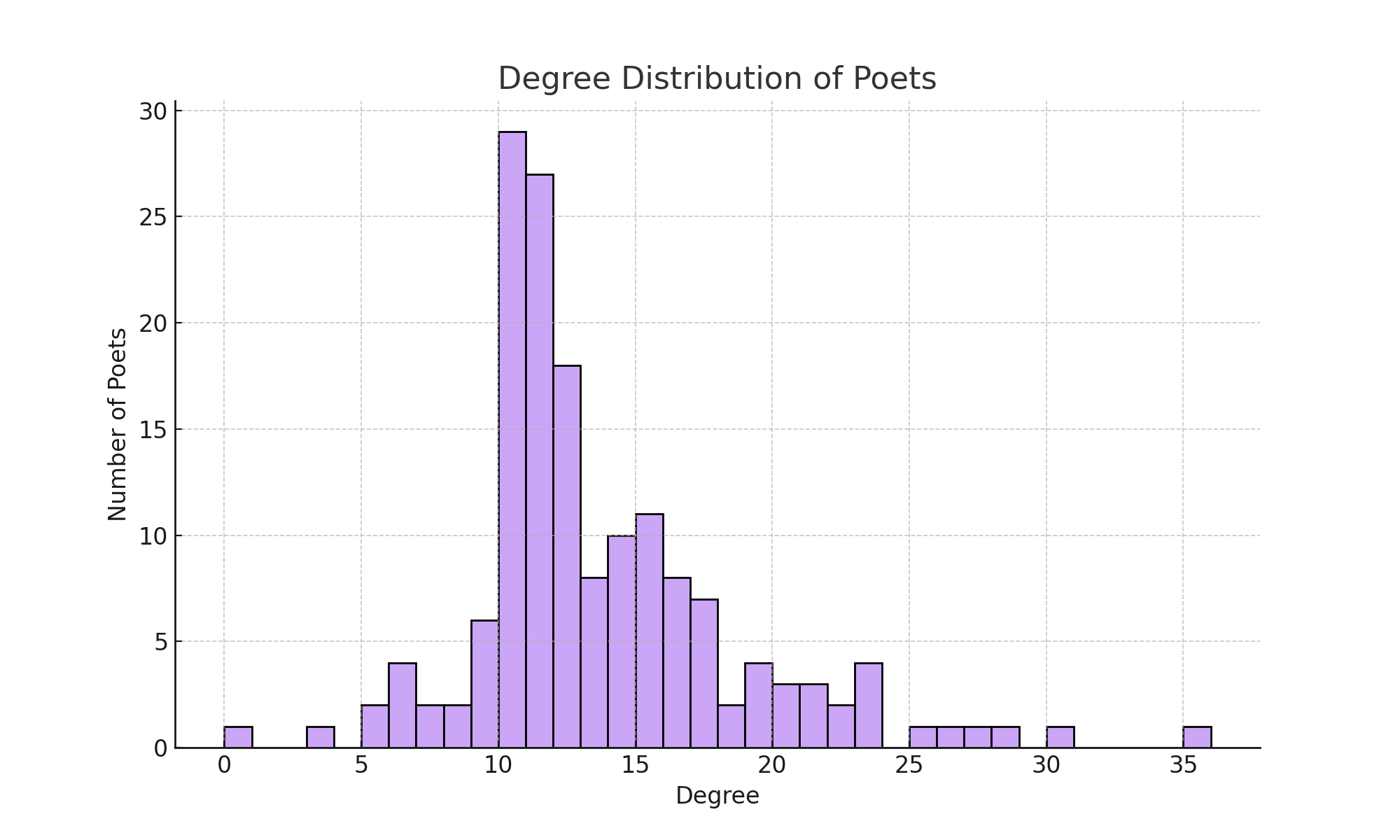}
    \caption{Degree distribution of poets in the similarity graph.}
    \label{fig:degree-distribution}
\end{figure}

Figure~\ref{fig:degree-distribution} shows the degree distribution of poets in the network. We observe a moderately right-skewed distribution, with a majority of poets having between 10 to 15 connections, but a small number exhibiting very high degrees.

\begin{itemize}
    \item Minimum degree: 0
    \item Maximum degree: 35
    \item Mean degree: 13.26
    \item Median degree: 12
    \item Standard deviation: 5.02
\end{itemize}

These figures show that although the network is rather well-connected generally, it is not homogeneous: some poets act as hubs in the similarity space, greatly more similar to others. Such poets deserve more study in centrality analysis and influence tracing since they may reflect stylistic archetypes or figures of great impact.

\subsection{Betweenness Centrality Distribution}

To further explore the structure of the poet similarity network, we examined the distribution of \textit{betweenness centrality} among all poets. Betweenness centrality measures how often a node (poet) lies on the shortest path between other pairs of nodes, thus indicating its role as a connector or bridge within the network.

\begin{figure}[H]
    \centering
    \includegraphics[width=0.7\textwidth]{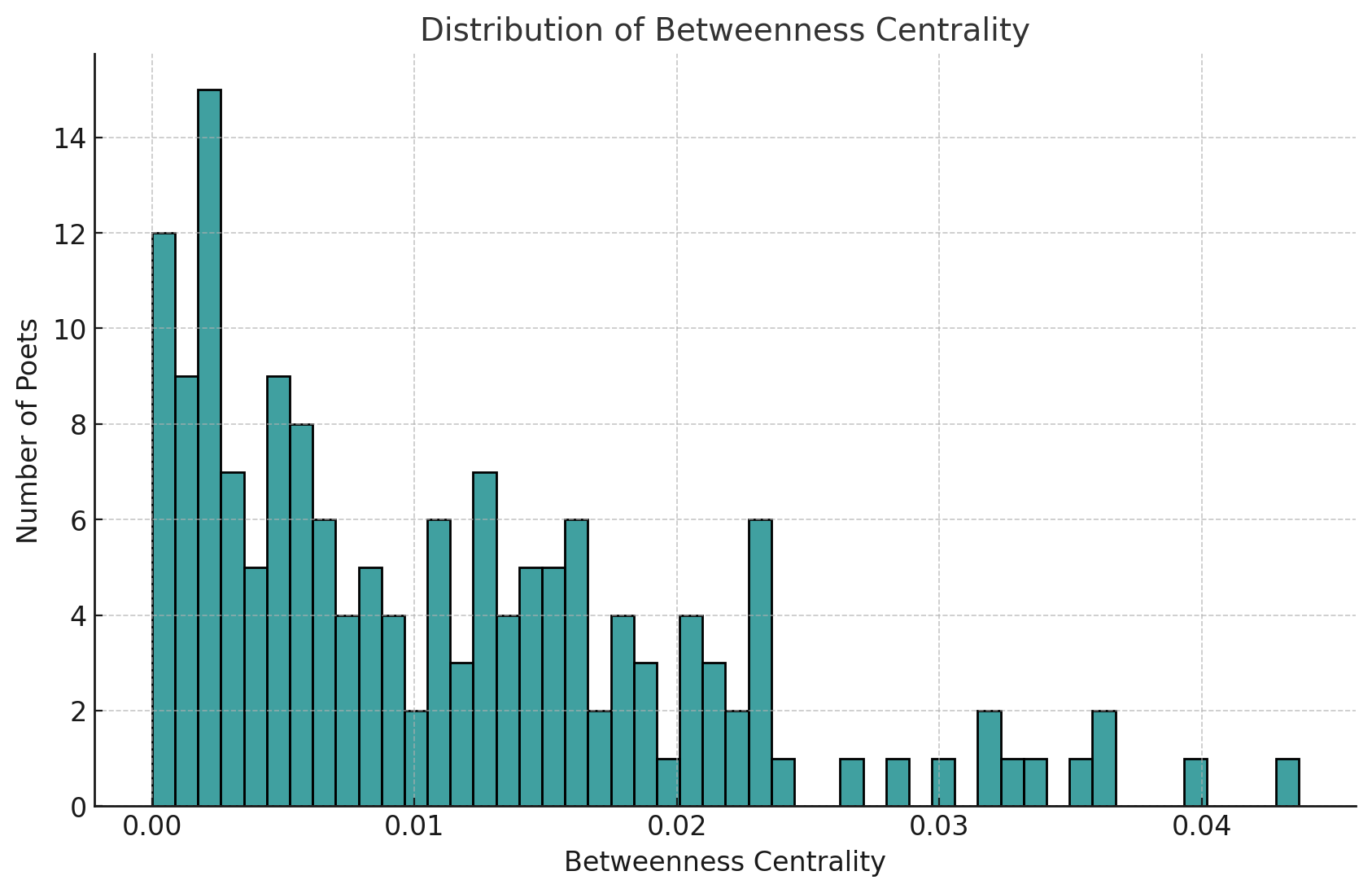}
    \caption{Distribution of Betweenness Centrality among Persian poets. Most poets lie near zero, while a few act as significant connectors in the network.}
    \label{fig:betweenness}
\end{figure}

Figure~\ref{fig:betweenness} shows the histogram of betweenness values. As seen, the distribution is highly right-skewed: a majority of poets have near-zero betweenness, while a small number serve as key intermediaries connecting different parts of the network. This is typical of complex networks with a modular structure, where only a few nodes facilitate cross-community interactions.

Such high-betweenness poets are critical for the flow of literary influence across stylistic and thematic boundaries. Their presence may suggest poets whose work spanned genres, time periods, or schools—thus positioning them centrally in the overall literary map.

\subsection{Eigenvector Centrality Distribution}

To better understand the prominence of poets in terms of their influence within the poetic network, we examine the distribution of \textbf{Eigenvector Centrality}. This measure assigns high scores to poets who are connected to other well-connected poets, thereby identifying central figures in the network’s overall structure.

\begin{figure}[H]
    \centering
    \includegraphics[width=0.7\textwidth]{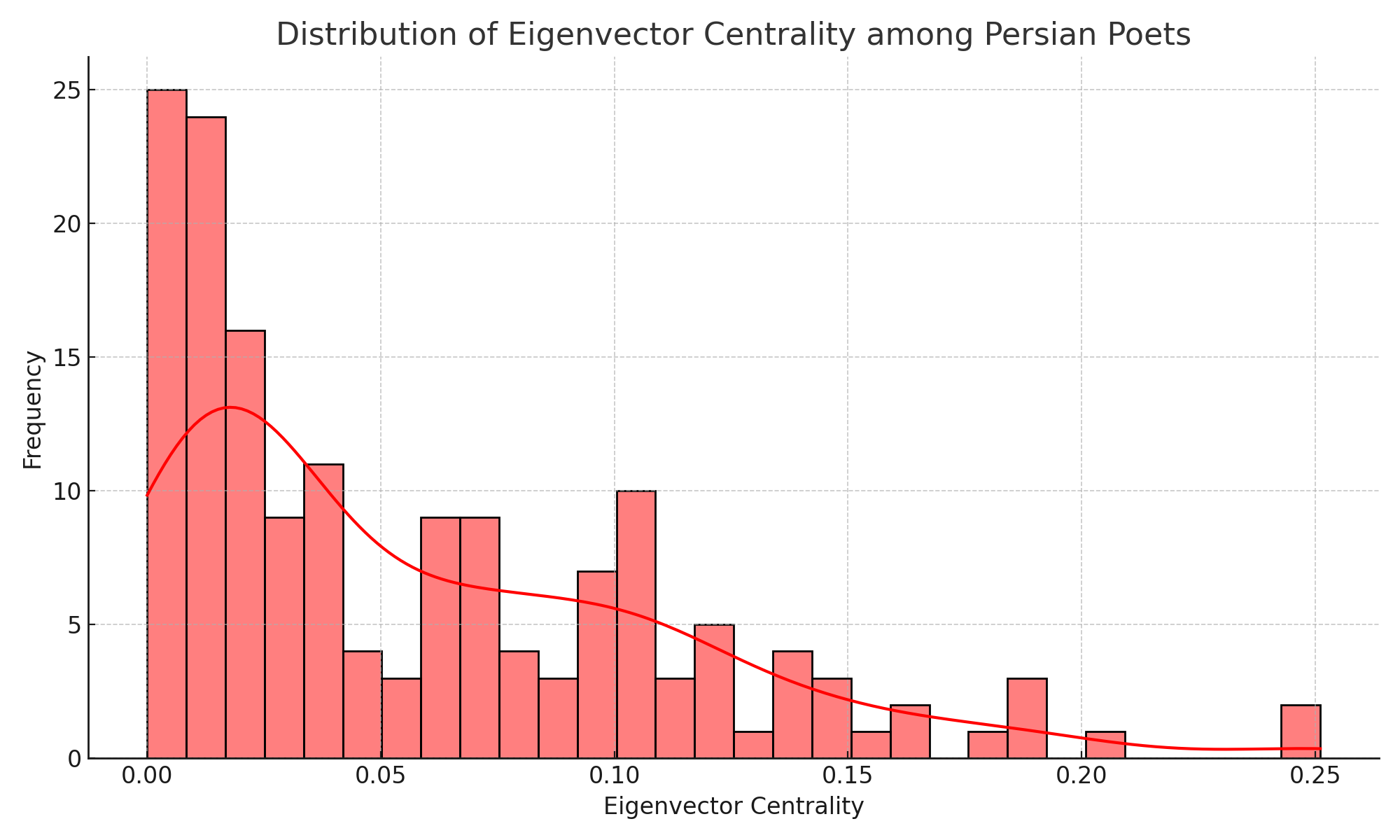}
    \caption{Distribution of Eigenvector Centrality across the poet influence network (red).}
    \label{fig:eigenvector_distribution}
\end{figure}

As shown in Figure~\ref{fig:eigenvector_distribution}, the distribution is highly skewed, with a small number of poets exhibiting significantly higher eigenvector centrality scores than the rest. This pattern suggests the existence of a poetic “elite” whose influence permeates the network, likely including canonical figures such as Rūmī, Hāfeẓ, Ferdowsī, and Saʿdī. The tail of the distribution flattens gradually, indicating a long list of poets with marginal centrality—reflecting either niche styles, thematic idiosyncrasy, or limited intertextual engagement.

Eigenvector centrality is particularly successful in underlining poets who are powerful not only because of their direct contacts but also because they are ingrained in rich clusters of mutual influence and respect. In Persian poetry, this can relate to fundamental writers who are referenced, copied, or stylistically echoed by others who are themselves powerful.

\subsection{Structural Properties of the Overall Graph}

We first study some basic graph-level statistics to better grasp the worldwide framework of the poet similarity network. There are 160 nodes total in the full graph—that of poets—and 1,061 edges—that of notable similarity relationships between them. This produces a network density of roughly 0.083, meaning that just roughly 8.3\% of all conceivable poet-to-poet connections exist. Typical for sparse real-world networks, this value indicates a structure far from random or totally connected.

\medskip
Interestingly, the graph is not fully connected; it comprises two connected components. However, the vast majority of poets fall within the largest component. For our structural analysis, we focus on this largest connected component (LCC), as it captures the primary network of interrelated poets.

\medskip
Within the LCC, the average shortest path length is approximately 2.78, indicating that any two poets in the main cluster can typically be reached via fewer than three steps. This reflects the \emph{small-world property} — a hallmark of many real-world networks — where most nodes can be reached from any other in a small number of steps.

\medskip
Furthermore, the network exhibits a relatively high average clustering coefficient of 0.479. This means that poets tend to form tightly interconnected local groups, consistent with our earlier observation of distinct literary communities. A clustering coefficient of nearly 0.5 implies that if poet A is similar to poet B and C, there is a high likelihood that B and C are also similar to each other — suggesting shared literary or stylistic lineage.

\medskip
These structural features together indicate that the poet similarity graph has both global cohesion and strong local clustering, making it well-suited for community detection and influence analysis.

\section{Conclusion}

Using network analysis, this work presented a computational framework for simulating influence and similarity in the traditional Persian poetry. We built a similarity graph of 160 canonical poets by using multi-dimensional textual elements covering semantic, lexical, stylistic, thematic, and metrical dimensions. Using the Louvain algorithm, community detection revealed eight unique poetic clusters that closely fit known literary styles and historical movements including the Sabk-e Hindī school, the Bazgasht-e Adabī revival, classical mystic lyricism, and the epic narrative tradition.

Through centrality analysis, we identified poets who act as hubs, bridges, or peripheral innovators within the literary network. Central figures such as Khāqānī, Asīr Akhsīkatī, and Qāʾānī emerged as influential across multiple metrics, while others like Ferdowsī, Rūmī, and ʿAṭṭār, though culturally paramount, occupied structurally peripheral roles—highlighting the distinction between literary fame and intertextual integration.

Along with recovering major stylistic flaws historically observed by literary experts, the resulting graph-based model also revealed surprising groups and latent nodes of influence. The structural elements of the network—such as its small-world character, strong community modularity, and hierarchical centrality patterns—confirm that classical Persian poetry consists in a densely linked but stylistically segmented tradition.

Our method shows that a strong lens for mapping poetic influence and tracking stylistic evolution is provided by network analysis when anchored in strong textual similarity measures. It links quantitative and humanistic approaches, so allowing fresh understanding of the dynamics of literary legacy. This work presents a scalable, interpretable model of classical Persian literature as a connected system of voices, styles, and ideas by formallyizing intertextual relations as measured links.

\section{Future Work}

The current study provides a foundational graph-based model of classical Persian poetic influence, yet it opens multiple avenues for further research and methodological refinement. Below, we outline several directions that could expand the scope and depth of this investigation:

\begin{itemize}
    \item \textbf{Temporal Dynamics:} While the present model operates on a static graph, future research could incorporate diachronic analysis by segmenting the corpus by century or poetic movement. This would enable temporal tracking of stylistic diffusion, network reconfiguration over time, and the rise or decline of poetic schools.


    \item \textbf{Citation and Paratextual Networks:} Integrating metadata such as explicit citations, marginalia, commentaries, and tazkiras could enrich the model with documentary evidence of influence, complementing the computational similarity-based edges with authorial and reader-based references.

    \item \textbf{Linguistic and Rhetorical Substructure Analysis:} While the study includes stylistic and thematic dimensions, deeper integration of syntactic, phonological, and rhetorical devices—such as enjambment, radif, or rhetorical figures—could uncover finer-grained patterns of stylistic imitation or divergence.

    \item \textbf{Cross-Cultural Comparison:} The methodology developed here could be adapted for other literary corpora, including Arabic, Ottoman Turkish, Urdu, or even European poetic traditions. Comparative network models would shed light on universal vs. culture-specific patterns of literary influence.

    \item \textbf{Interactive Visualization and Public Humanities Tools:} Building a web-based platform for exploring the Persian poets network interactively would make this work accessible to broader audiences, including educators, students, and the public. Dynamic graph visualizations, cluster exploration, and timeline overlays could enrich cultural understanding.

    \item \textbf{Generative Modeling:} Finally, embedding the graph in a generative framework—such as graph neural networks or temporal embedding models—may allow for the prediction of plausible stylistic links or the simulation of hypothetical influence pathways across literary history.
\end{itemize}

Although much more research has to be done, overall the present model provides a strong basis for the computational study of Persian literary tradition. Unlocking the full possibilities of literary network analysis depends on further developments in natural language processing, historical corpus development, and multidisciplinary cooperation.

\section{Limitations}

Despite the breadth and depth of this study, several limitations should be acknowledged to contextualize the findings and frame future improvements:

\begin{itemize}
    \item \textbf{Corpus Coverage:} The dataset used in this study, while extensive, is limited to poets represented in the Ganjoor corpus. This excludes a number of lesser-known poets, regional figures, or recent literary voices whose inclusion might influence network topology.

    \item \textbf{Disputed Authorship and Attribution Errors:} Some poems in the corpus may have disputed or uncertain attribution. While the study attempted to address this by marking and excluding ambiguous cases, misattributions could still affect similarity scores and community assignments.

    \item \textbf{Edge Thresholding and Graph Construction:} The final graph was constructed by thresholding similarity values. While this method yields a manageable and interpretable network, it introduces arbitrariness and may omit weaker but meaningful connections.

    \item \textbf{Semantic and Stylistic Feature Limits:} Although the analysis combined multiple feature types, some dimensions—such as rhetorical figures, dialectal variation, or audience reception—were not included. These omissions may affect the completeness of the similarity model.

    \item \textbf{Language Model Constraints:} The use of pre-trained multilingual and Persian NLP models (e.g., sentence transformers, POS taggers) introduces dependency on existing model quality, which may not fully capture the nuance of classical Persian poetry, especially archaic or esoteric vocabulary.

    \item \textbf{Network Interpretability:} Network metrics such as centrality or modularity offer structural insights, but their literary interpretation remains partly subjective and requires careful historical contextualization, which may vary depending on scholarly perspective.
\end{itemize}

Acknowledging these constraints helps us to understand the results with suitable care while appreciating the significant advancement this work shows in connecting computational modeling with literary analysis.

\section*{Acknowledgments}

The authors would want to thank the developers and curators of the Ganjoor Project, whose open-access corpus of Persian poetry provided the fundamental dataset for this study. We also thank the creators of the open-source NLP tools—especially the Hazm toolkit and pretrained Persian embeddings—without which major computational analysis of Persian literary texts would be much more limited.

Talks with colleagues in both literary studies and data science benefited this work since their multidisciplinary insights shaped the network modeling approach. We much appreciate the help of open digital infrastructure communities as well as academic ones.

\vspace{6pt}

\informedconsent{Informed consent was obtained from all subjects involved in the study.}

\dataavailability{}

\conflictsofinterest{The authors declare no conflicts of interest.} 



\reftitle{References}

\bibliography{bibliography.bib}

\isPreprints{}{}

\newpage

\appendix
\section*{Appendix A: Poets Combined Centrality Measures}
\begin{center}
\renewcommand{\arraystretch}{1.2}
\begin{longtable}{lccccc}
\caption*{Poets Combined Centrality Measures (Raw Values, Sorted by Normalized Total)}\\
\toprule
\textbf{Poet} & \textbf{Degree} & \textbf{Betweenness} & \textbf{Closeness} & \textbf{Eigenvector} & \textbf{Katz} \\
\midrule
\endfirsthead
\multicolumn{6}{c}{\textit{(continued from previous page)}}\\
\toprule
\textbf{Poet} & \textbf{Degree} & \textbf{Betweenness} & \textbf{Closeness} & \textbf{Eigenvector} & \textbf{Katz} \\
\midrule
\endhead
\bottomrule
\endfoot
Khāqānī & 0.2201 & 0.0437 & 0.4176 & 0.2499 & 0.0877 \\
Athīr Akhsīkatī & 0.1887 & 0.0228 & 0.3945 & 0.2511 & 0.0859 \\
Qāʾānī & 0.1698 & 0.0365 & 0.3935 & 0.1857 & 0.0845 \\
Ṣāmet Borūjerdī & 0.1635 & 0.0193 & 0.3995 & 0.2074 & 0.0842 \\
Khāled Naqshbandī & 0.1572 & 0.0203 & 0.3886 & 0.1864 & 0.0838 \\
Jahān Malek Khātūn & 0.1761 & 0.0361 & 0.3915 & 0.0314 & 0.0849 \\
Majd-e Hamgar & 0.1384 & 0.0217 & 0.4099 & 0.1641 & 0.0826 \\
Adīb al-Mamālik & 0.1384 & 0.0231 & 0.3886 & 0.1659 & 0.0826 \\
Ḥakīm Sabzevārī & 0.1447 & 0.0232 & 0.3896 & 0.1449 & 0.0829 \\
Adīb Ṣābir & 0.1447 & 0.0188 & 0.3955 & 0.1560 & 0.0830 \\
Afsar Kirmānī & 0.1447 & 0.0128 & 0.3848 & 0.1872 & 0.0830 \\
Salmān Sāvojī & 0.1195 & 0.0321 & 0.4078 & 0.0992 & 0.0813 \\
Ḥakīm Nizārī & 0.1447 & 0.0325 & 0.3792 & 0.0227 & 0.0829 \\
Moḥīṭ Qommī & 0.1321 & 0.0093 & 0.3792 & 0.1769 & 0.0822 \\
Amīr Muʿizzī & 0.1258 & 0.0129 & 0.3839 & 0.1350 & 0.0818 \\
Naẓīrī Neyshābūrī & 0.1258 & 0.0146 & 0.3820 & 0.1203 & 0.0817 \\
Āzar Bīgdelī & 0.0755 & 0.0397 & 0.4143 & 0.0492 & 0.0785 \\
Ahlī Shīrāzī & 0.1069 & 0.0232 & 0.3965 & 0.0666 & 0.0805 \\
Ḥeydar Shīrāzī & 0.1132 & 0.0075 & 0.3792 & 0.1491 & 0.0810 \\
Rażī al-Dīn Ārtīmānī & 0.1258 & 0.0264 & 0.3738 & 0.0178 & 0.0817 \\
Vaḥdat Kermānshāhī & 0.1069 & 0.0111 & 0.3774 & 0.1369 & 0.0806 \\
Amīr Shāhī & 0.1195 & 0.0229 & 0.3877 & 0.0348 & 0.0813 \\
Homām Tabrīzī & 0.0943 & 0.0336 & 0.4057 & 0.0195 & 0.0796 \\
Rūdakī & 0.1006 & 0.0321 & 0.3765 & 0.0255 & 0.0801 \\
Nasīmī & 0.1069 & 0.0139 & 0.3802 & 0.1124 & 0.0806 \\
Jalāl ʿAżud & 0.1321 & 0.0203 & 0.3660 & 0.0274 & 0.0820 \\
Ẓahīr Fāryābī & 0.1132 & 0.0061 & 0.3626 & 0.1468 & 0.0810 \\
Bābā Fag͟hānī & 0.1006 & 0.0188 & 0.3802 & 0.0865 & 0.0801 \\
Ibn Ḥesām Khosfī & 0.1006 & 0.0113 & 0.3783 & 0.1248 & 0.0802 \\
Jūyā-ye Tabrīzī & 0.1006 & 0.0136 & 0.3634 & 0.1175 & 0.0802 \\
Malek al-Shoʿarāʾ Bahār & 0.0818 & 0.0227 & 0.3820 & 0.0941 & 0.0790 \\
Masʿūd Saʿd Salmān & 0.1069 & 0.0176 & 0.3820 & 0.0640 & 0.0805 \\
Jāmī & 0.0943 & 0.0133 & 0.3906 & 0.1091 & 0.0798 \\
Parvīn Eʿteṣāmī & 0.1321 & 0.0150 & 0.3721 & 0.0227 & 0.0821 \\
Saḥāb-e Eṣfahānī & 0.0943 & 0.0131 & 0.3886 & 0.1074 & 0.0798 \\
Ḥazīn Lāhījī & 0.1069 & 0.0077 & 0.3703 & 0.1141 & 0.0806 \\
Mollā Masīḥ & 0.0692 & 0.0354 & 0.3792 & 0.0367 & 0.0780 \\
Qaṭrān Tabrīzī & 0.1006 & 0.0144 & 0.3738 & 0.0739 & 0.0801 \\
Reżāqolī Khān Hedāyat & 0.1006 & 0.0089 & 0.3774 & 0.1010 & 0.0801 \\
Ṣafā-ye Eṣfahānī & 0.0943 & 0.0089 & 0.3738 & 0.1024 & 0.0798 \\
Jamāl al-Dīn ʿAbd al-Razzāq & 0.0943 & 0.0155 & 0.3783 & 0.0625 & 0.0797 \\
Mīrzādeh ʿEshqī & 0.1069 & 0.0156 & 0.3915 & 0.0242 & 0.0806 \\
Khwāju-ye Kirmānī & 0.0692 & 0.0301 & 0.3601 & 0.0466 & 0.0781 \\
Mujīr al-Dīn Baylaqānī & 0.0881 & 0.0080 & 0.3774 & 0.1177 & 0.0794 \\
Fożūlī & 0.0943 & 0.0163 & 0.3802 & 0.0545 & 0.0797 \\
Gharavī Eṣfahānī & 0.0943 & 0.0023 & 0.3585 & 0.1373 & 0.0799 \\
Neshāṭ Eṣfahānī & 0.1069 & 0.0178 & 0.3668 & 0.0175 & 0.0805 \\
Vaṭvāṭ & 0.0943 & 0.0025 & 0.3544 & 0.1378 & 0.0799 \\
Qāsem Anvār & 0.0881 & 0.0187 & 0.3820 & 0.0449 & 0.0793 \\
Saʿd al-Dīn Varāvīnī & 0.0818 & 0.0175 & 0.3765 & 0.0662 & 0.0789 \\
Sayyid Ḥasan G͟haznavī & 0.1006 & 0.0099 & 0.3643 & 0.0720 & 0.0801 \\
Sayf Farghānī & 0.0818 & 0.0229 & 0.3802 & 0.0286 & 0.0790 \\
Amīr Khusrow Dehlavī & 0.1195 & 0.0112 & 0.3536 & 0.0223 & 0.0813 \\
Wāʿeẓ Qazvīnī & 0.0881 & 0.0055 & 0.3839 & 0.1066 & 0.0794 \\
Moḥtasham Kāshānī & 0.0692 & 0.0207 & 0.3995 & 0.0511 & 0.0781 \\
Saʿdī & 0.0881 & 0.0223 & 0.3721 & 0.0115 & 0.0793 \\
ʿAsjodī & 0.1006 & 0.0063 & 0.3536 & 0.0845 & 0.0801 \\
Mawlānā & 0.0943 & 0.0178 & 0.3756 & 0.0164 & 0.0797 \\
Azraqī Haravī & 0.0881 & 0.0073 & 0.3489 & 0.1042 & 0.0794 \\
Ibn Yamīn & 0.0755 & 0.0125 & 0.3829 & 0.0764 & 0.0786 \\
Ḥosayn Khwārazmī & 0.0692 & 0.0163 & 0.3965 & 0.0608 & 0.0781 \\
ʿAmaq Bukhārī & 0.0881 & 0.0068 & 0.3618 & 0.0918 & 0.0794 \\
ʿĀref Qazvīnī & 0.0755 & 0.0214 & 0.3820 & 0.0209 & 0.0785 \\
Halālī Jughṭāʾī & 0.1195 & 0.0060 & 0.3601 & 0.0212 & 0.0813 \\
Fayż Kāshānī & 0.0943 & 0.0163 & 0.3643 & 0.0171 & 0.0797 \\
ʿAyn al-Qożāt Hamadānī & 0.0692 & 0.0287 & 0.3489 & 0.0135 & 0.0780 \\
Amīr ʿAlīshīr Nawāʾī & 0.0881 & 0.0019 & 0.3451 & 0.1180 & 0.0794 \\
Mollā Aḥmad Narāqī & 0.0629 & 0.0243 & 0.3811 & 0.0270 & 0.0777 \\
Salīmī Jerūnī & 0.0881 & 0.0173 & 0.3643 & 0.0163 & 0.0793 \\
Ghobār Hamadānī & 0.0755 & 0.0107 & 0.3765 & 0.0754 & 0.0785 \\
Ḥamīd al-Dīn Balkhī & 0.0818 & 0.0057 & 0.3677 & 0.0940 & 0.0790 \\
Falakī Sharvānī & 0.0818 & 0.0011 & 0.3520 & 0.1266 & 0.0791 \\
Abū’l-Faraj Rūnī & 0.0755 & 0.0050 & 0.3703 & 0.1065 & 0.0786 \\
Anvarī & 0.0755 & 0.0136 & 0.3677 & 0.0586 & 0.0785 \\
Abū’l-Ḥasan Farāhānī & 0.0943 & 0.0123 & 0.3651 & 0.0162 & 0.0797 \\
Rashḥa & 0.0755 & 0.0081 & 0.3703 & 0.0737 & 0.0786 \\
Yaghmā-ye Jandaqī & 0.0755 & 0.0031 & 0.3643 & 0.1048 & 0.0786 \\
Shahriyār & 0.0755 & 0.0037 & 0.3593 & 0.1030 & 0.0786 \\
Naṣrullāh Monshī & 0.0755 & 0.0181 & 0.3738 & 0.0133 & 0.0784 \\
Meybodī & 0.0566 & 0.0161 & 0.3601 & 0.0703 & 0.0774 \\
Salīm Tehrānī & 0.0692 & 0.0084 & 0.3756 & 0.0727 & 0.0782 \\
ʿOmān Sāmānī & 0.0692 & 0.0055 & 0.3634 & 0.0952 & 0.0782 \\
Qaṣṣāb Kāshānī & 0.0755 & 0.0160 & 0.3756 & 0.0143 & 0.0785 \\
Sanāʾī & 0.0692 & 0.0119 & 0.3811 & 0.0442 & 0.0781 \\
ʿAyyūqī & 0.0692 & 0.0216 & 0.3481 & 0.0086 & 0.0781 \\
Elhāmī Kermānshāhī & 0.0692 & 0.0203 & 0.3512 & 0.0121 & 0.0780 \\
Solṭān Walad & 0.0692 & 0.0159 & 0.3811 & 0.0180 & 0.0781 \\
Ṭabīb Eṣfahānī & 0.0692 & 0.0155 & 0.3756 & 0.0191 & 0.0782 \\
Hāfeẓ & 0.0755 & 0.0121 & 0.3552 & 0.0381 & 0.0785 \\
Shāhedī & 0.0692 & 0.0125 & 0.3774 & 0.0348 & 0.0782 \\
Qāʾem Maqām Farāhānī & 0.0692 & 0.0054 & 0.3651 & 0.0804 & 0.0782 \\
Najm al-Dīn Rāzī & 0.0629 & 0.0109 & 0.3552 & 0.0672 & 0.0778 \\
Sheykh Bahāʾī & 0.0818 & 0.0120 & 0.3626 & 0.0112 & 0.0789 \\
Kalīm & 0.0692 & 0.0148 & 0.3660 & 0.0206 & 0.0780 \\
Ghāleb Dehlavī & 0.0818 & 0.0053 & 0.3489 & 0.0546 & 0.0789 \\
Qodsī Mashhadī & 0.0629 & 0.0144 & 0.3765 & 0.0237 & 0.0777 \\
Emāmī Haravī & 0.0692 & 0.0010 & 0.3443 & 0.1034 & 0.0782 \\
Manūchihrī & 0.0818 & 0.0017 & 0.3348 & 0.0709 & 0.0789 \\
Niẓāmī & 0.0755 & 0.0143 & 0.3436 & 0.0081 & 0.0785 \\
Owhad al-Dīn Kermānī & 0.0881 & 0.0101 & 0.3391 & 0.0072 & 0.0792 \\
Nayyer Tabrīzī & 0.0692 & 0.0008 & 0.3355 & 0.0935 & 0.0782 \\
Ṭoġrul Aḥrārī & 0.0629 & 0.0009 & 0.3552 & 0.0940 & 0.0778 \\
Vaḥīd al-Zamān Qazvīnī & 0.0629 & 0.0006 & 0.3443 & 0.1020 & 0.0779 \\
Noʿī Khabūshānī & 0.0629 & 0.0048 & 0.3552 & 0.0706 & 0.0778 \\
ʿAbd al-Vāseʿ Jabalī & 0.0629 & 0.0027 & 0.3544 & 0.0793 & 0.0778 \\
Khayālī Bukhārāʾī & 0.0755 & 0.0062 & 0.3634 & 0.0274 & 0.0785 \\
ʿObeyd Zākānī & 0.0629 & 0.0051 & 0.3544 & 0.0664 & 0.0777 \\
Farrokhī Yazdī & 0.0692 & 0.0150 & 0.3376 & 0.0069 & 0.0780 \\
Asīrī Lāhījī & 0.0629 & 0.0050 & 0.3634 & 0.0608 & 0.0777 \\
Hātif Iṣfahānī & 0.0629 & 0.0005 & 0.3443 & 0.0958 & 0.0778 \\
ʿOrfī & 0.0692 & 0.0021 & 0.3428 & 0.0683 & 0.0782 \\
Iraj Mīrzā & 0.0881 & 0.0043 & 0.3391 & 0.0151 & 0.0793 \\
Shāh Niʿmatullāh Walī & 0.0566 & 0.0125 & 0.3811 & 0.0138 & 0.0773 \\
Iqbal Lāhorī & 0.0629 & 0.0029 & 0.3585 & 0.0666 & 0.0777 \\
Kamāl al-Dīn Esmāʿīl & 0.0692 & 0.0077 & 0.3481 & 0.0316 & 0.0781 \\
Asīr-e Shahrastānī & 0.0692 & 0.0063 & 0.3466 & 0.0376 & 0.0781 \\
Niẓāmī ʿArūżī & 0.0629 & 0.0113 & 0.3568 & 0.0112 & 0.0777 \\
Āshefta-ye Shīrāzī & 0.0629 & 0.0079 & 0.3576 & 0.0232 & 0.0777 \\
Saʿīdā & 0.0692 & 0.0051 & 0.3576 & 0.0147 & 0.0782 \\
Farrokhī Sīstānī & 0.0692 & 0.0028 & 0.3413 & 0.0380 & 0.0781 \\
Forūghī Bastāmī & 0.0566 & 0.0065 & 0.3668 & 0.0259 & 0.0773 \\
Kasāʾī & 0.0629 & 0.0049 & 0.3384 & 0.0370 & 0.0777 \\
Shāṭer ʿAbbās Ṣobūḥī & 0.0629 & 0.0034 & 0.3552 & 0.0342 & 0.0777 \\
Īrānshān & 0.0755 & 0.0057 & 0.3292 & 0.0061 & 0.0784 \\
ʿOnsor al-Maʿālī & 0.0755 & 0.0050 & 0.3264 & 0.0064 & 0.0784 \\
Abū Saʿīd Abū’l-Khayr & 0.0755 & 0.0047 & 0.3292 & 0.0054 & 0.0784 \\
Ṣāʾeb Tabrīzī & 0.0629 & 0.0022 & 0.3466 & 0.0354 & 0.0777 \\
Fakhr al-Dīn Asʿad Gorgānī & 0.0566 & 0.0093 & 0.3436 & 0.0100 & 0.0773 \\
ʿIrāqī & 0.0755 & 0.0023 & 0.3319 & 0.0127 & 0.0785 \\
Kamāl Khojandī & 0.0692 & 0.0027 & 0.3319 & 0.0148 & 0.0781 \\
Ferdowsī & 0.0692 & 0.0043 & 0.3264 & 0.0068 & 0.0780 \\
Mahastī Ganjavī & 0.0566 & 0.0082 & 0.3285 & 0.0075 & 0.0773 \\
Afsar al-Molūk ʿĀmelī & 0.0692 & 0.0034 & 0.3285 & 0.0058 & 0.0781 \\
Bābā Afżal Kāshānī & 0.0629 & 0.0065 & 0.3172 & 0.0078 & 0.0777 \\
ʿOnsorī & 0.0629 & 0.0005 & 0.3128 & 0.0403 & 0.0777 \\
Nāṣir-i Khusraw & 0.0629 & 0.0015 & 0.3391 & 0.0151 & 0.0778 \\
ʿAbd al-Qādir Gīlānī & 0.0629 & 0.0009 & 0.3443 & 0.0129 & 0.0778 \\
Owhadī & 0.0629 & 0.0012 & 0.3384 & 0.0131 & 0.0778 \\
Mīrzā Āqā Khān Kermānī & 0.0629 & 0.0041 & 0.3257 & 0.0066 & 0.0776 \\
Solṭān Bāhū & 0.0629 & 0.0007 & 0.3376 & 0.0123 & 0.0778 \\
Ghazālī & 0.0692 & 0.0026 & 0.3079 & 0.0051 & 0.0780 \\
Shams Maghribī & 0.0629 & 0.0007 & 0.3237 & 0.0119 & 0.0778 \\
Moḥammad Kawsaj & 0.0629 & 0.0019 & 0.3165 & 0.0052 & 0.0777 \\
Moḥammad b. Monavvar & 0.0629 & 0.0024 & 0.3109 & 0.0051 & 0.0777 \\
Bīdil Dehlavī & 0.0503 & 0.0009 & 0.3333 & 0.0263 & 0.0769 \\
Khājeh ʿAbdullāh Anṣārī & 0.0629 & 0.0018 & 0.2979 & 0.0039 & 0.0776 \\
Bābā Ṭāher & 0.0377 & 0.0011 & 0.3398 & 0.0363 & 0.0761 \\
Khayyām & 0.0629 & 0.0014 & 0.2908 & 0.0036 & 0.0776 \\
ʿOsmān Mokhtārī & 0.0629 & 0.0005 & 0.2991 & 0.0041 & 0.0776 \\
ʿAṭṭār & 0.0503 & 0.0020 & 0.3231 & 0.0077 & 0.0769 \\
Vahshī Bāfaqī & 0.0440 & 0.0021 & 0.3348 & 0.0130 & 0.0765 \\
Rahi Moʿayyeri & 0.0377 & 0.0042 & 0.3376 & 0.0071 & 0.0761 \\
Moshtāq Eṣfahānī & 0.0314 & 0.0045 & 0.3497 & 0.0118 & 0.0757 \\
Khājeh Naṣīr al-Dīn Ṭūsī & 0.0314 & 0.0022 & 0.3391 & 0.0243 & 0.0757 \\
Author of Farāmarznāmeh & 0.0566 & 0.0001 & 0.2897 & 0.0037 & 0.0772 \\
Asadī Ṭūsī & 0.0377 & 0.0021 & 0.3134 & 0.0046 & 0.0761 \\
Khalīlullāh Khalīlī & 0.0440 & 0.0007 & 0.2979 & 0.0028 & 0.0765 \\
Daqīqī & 0.0377 & 0.0021 & 0.2991 & 0.0051 & 0.0761 \\
Fāyez & 0.0189 & 0.0006 & 0.2985 & 0.0032 & 0.0749 \\
Sheykh Maḥmūd Shabistarī & 0.0000 & 0.0000 & 0.0000 & 0.0000 & 0.0737 \\
\end{longtable}
\end{center}
\label{tab:appendix_poet_centralities}

\newpage

\appendix
\section*{Appendix B: Louvain Community Detection Results}

\appendix
\subsection*{Community 1 (18 poets)}
\begin{quote}
  Mollā Masīḥ, Salmān Sāvojī, Āzar Bīgdelī, Amīr Shāhī, Ahlī Shīrāzī, Jāmī, Ḥāfeẓ, Khiyālī Bukhārāʾī, Rashḥa, Saḥāb Iṣfahānī, Salīm Tehrānī, Shāṭer ʿAbbās Ṣobūḥī, Shāhedī, Qāsem Anvār, Qodsī Mashhadī, Nasīmī, Wāʿeẓ Qazvīnī, Vaḥshī Bāfqī
\end{quote}

\subsection*{Community 2 (19 poets)}
\begin{quote}
  Homām Tabrīzī, Āshefta-ye Shīrāzī, Ebn-e Yamīn, Asīr-e Shahrestānī, Bābā Faqānī, Bīdel Dehlavī, Ḥazīn Lāhījī, Ḥoseyn Khwārazmī, Rahī Moʿeīrī, Shāh Neʿmatollāh Valī, Ṣāʾeb Tabrīzī, ʿOrfī, Ghāleb Dehlavī, Forūghī Basṭāmī, Fożūlī, Moḥtashem Kāshānī, Moshtāq Eṣfahānī, Naẓīrī Neyshābūrī, Kalīm
\end{quote}

\subsection*{Community 3 (18 poets)}
\begin{quote}
  Abū Saʿīd Abū’l-Khayr, Owḥad al-Dīn Kermānī, Bābā Afżal Kāshānī, Khalīlallāh Khalīlī, Khwāja ʿAbdullāh Anṣārī, Khayyām, Saʿdī, Sheykh Bahāʾī, ʿĀref Qazvīnī, ʿOnsor al-Maʿālī, ʿAyn al-Qożāt Hamadānī, Ghazālī, Farrukhī Yazdī, Moḥammad b. Monawwar, Mollā Aḥmad Narāqī, Mahastī Ganjavī, Naṣrallāh Monshī, Neẓāmī ʿArowżī
\end{quote}

\subsection*{Community 4 (31 poets)}
\begin{quote}
  Jahān Malek Khātūn, Abū al-Ḥasan Farāhānī, Amīr Khosrow Dehlavī, Owḥadī, Īraj Mīrzā, Jalāl ʿAżod, Ḥakīm Nazarī, Daqīqī, Rażī al-Dīn Ārtīmānī, Rūdakī, Saʿīdā, Solṭān Bāhū, Solṭān Walad, Salīmī Jarūnī, Sayf Farghānī, Shams-e Moghrebī, Ṭabīb Eṣfahānī, ʿAbd al-Qāder Gīlānī, ʿIrāqī, ʿAṭṭār, Fāyez, Fakhr al-Dīn Asʿad Gorgānī, Fayż Kāshānī, Qaṣṣāb Kāshānī, Mawlānā, Mīrzādeh ʿEshqī, Nāṣer-e Khosrow, Neshāṭ Eṣfahānī, Helālī Joghṭāʾī, Parvīn Eʿteṣāmī, Kamāl Khojandī
\end{quote}

\subsection*{Community 5 (1 poets)}
\begin{quote}
  Sheykh Maḥmūd Shabestarī
\end{quote}

\subsection*{Community 6 (17 poets)}
\begin{quote}
  Adīb Ṣāber, Azraqī Haravī, Amīr Moʿezzī, Anwarī, Jamāl al-Dīn ʿAbd al-Razzāq, Sanāʾī, Sayyed Ḥasan Ghaznavī, ʿAbd al-Wāseʿ Jabalī, ʿAsjadī, ʿAmaq-e Bokhārī, ʿOnsorī, Farrukhī Sīstānī, Qaṭrān Tabrīzī, Masʿūd Saʿd Salmān, Manūchehrī, Kasāʾī, Kamāl al-Dīn Esmāʿīl
\end{quote}

\subsection*{Community 7 (12 poets)}
\begin{quote}
  Asadī Ṭūsī, Afsar al-Molūk ʿĀmelī, Elhāmī Kermānshāhī, Īrānshān, Khwājū-ye Kermānī, Sarayande-ye Farāmarznāmeh, ʿOthmān Mokhtārī, ʿAyyūqī, Ferdowsī, Moḥammad Kowsaj, Mīrzā Āqā Khān Kirmānī, Neẓāmī
\end{quote}

\subsection*{Community 8 (44 poets)}
\begin{quote}
  Afṣar Kirmānī, Asīr Akhsīkatī, Khāled Naqshbandī, Khāqānī, Majd-e Hamgar, Qāʾānī, Ebn-e Ḥosām Khūsfī, Abū al-Faraj Rūnī, Adīb al-Mamālek, Asīrī Lāhījī, Eqbāl Lāhorī, Emāmī Haravī, Amīr ʿAlīshīr Navāʾī, Bābā Ṭāher, Jūyā-ye Tabrīzī, Ḥamīd al-Dīn Balkhī, Ḥakīm Sabzawārī, Ḥeydar Shīrāzī, Khwāja Naṣīr al-Dīn Ṭūsī, Reżāqolī Khān Hedāyat, Saʿd al-Dīn Varāwīnī, Shahryār, Ṣafā-ye Eṣfahānī, Ṭoghrul Aḥrārī, Ẓahīr Fāryābī, ʿObayd Zākānī, ʿOmmān Sāmānī, Ghobār-e Hamadānī, Gharavī Eṣfahānī, Falakī Sharwānī, Qāʾem Maqām Farāhānī, Mojīr al-Dīn Baylaqānī, Moḥīṭ Qommī, Malek al-Shoʿarāʾ Bahār, Meybodī, Najm al-Dīn Rāzī, Nowʿī Khabūshānī, Nayyer Tabrīzī, Hātef Eṣfahānī, Vaḥdat Kermānshāhī, Vaḥīd al-Zamān Qazvīnī, Vaṭvāṭ, Yaghmā-ye Jandaqī, Ṣāmet Borūjerdī
\end{quote}

\end{document}